\documentclass[11pt,a4paper]{article}
\pdfoutput=1 
\usepackage{jheppub}

\usepackage{ifpdf}
\usepackage{afterpage}
\usepackage{amssymb}
\usepackage{amsmath}
\usepackage{graphicx}
\usepackage{longtable}
\usepackage{verbatim}
\usepackage{amsfonts}
\usepackage{bbold}
\usepackage[table,usenames,dvipsnames]{xcolor}
\usepackage{titlesec}
\usepackage{enumerate}
\usepackage[multiple]{footmisc}

\usepackage{physics} 
\usepackage{subcaption}

\usepackage{tikz}
\usetikzlibrary{trees}
\usetikzlibrary{decorations.pathmorphing}
\usetikzlibrary{decorations.markings}
\usetikzlibrary{decorations.pathreplacing}
\tikzset{
    photon/.style={decorate, decoration={snake}, draw=black},
    wino/.style={draw=redwine},
    electron/.style={draw=black, postaction={decorate},
        decoration={markings,mark=at position .55 with {\arrow[draw=black]{>}}}},
    scalar/.style={draw=black, dashed,postaction={decorate},
        decoration={markings,mark=at position .55 with {\arrow[draw=black]{>}}}},
    gluon/.style={decorate, draw=black,
        decoration={coil,amplitude=4pt, segment length=5pt}}
}

\newcommand{\bear}{\begin{array}}
\newcommand{\ear}{\end{array}}
\newcommand{\beq}{\begin{equation}}
\newcommand{\eeq}{\end{equation}}
\newcommand{\beqa}{\begin{eqnarray}}
\newcommand{\eeqa}{\end{eqnarray}}

\def\OMIT#1{{}}
\newcommand{\lsim}{\mathrel{\rlap{\lower4pt\hbox{\hskip1pt$\sim$}}
    \raise1pt\hbox{$<$}}}         
\newcommand{\gsim}{\mathrel{\rlap{\lower4pt\hbox{\hskip1pt$\sim$}}
    \raise1pt\hbox{$>$}}}         

\newcommand{\Sec}[1]{Sec.~\ref{#1}}

\newcommand{\Fig}[1]{Fig.~\ref{#1}}
\newcommand{\Eq}[1]{Eq.~(\ref{#1})}

\newcommand{\ignore}[1]{}

\newcommand{\MP}{M_{\mathrm{pl}}}

\vspace*{-30mm}

\title{\boldmath Spillway Preheating}
\author[a]{JiJi Fan,}
\author[b]{Kaloian D. Lozanov,}
\author[c]{Qianshu Lu}
\affiliation[a]{Department of Physics \& Brown Theoretical Physics Center, Brown University,
Providence, RI, 02912, USA}
\affiliation[b]{Illinois Center for Advanced Studies of the Universe \& Department of Physics, University of Illinois at Urbana-Champaign, Urbana, IL 61801, USA}
\affiliation[c]{Department of Physics, Harvard University, Cambridge, MA, 02138. USA}
\emailAdd{jiji\_fan@brown.edu}
\emailAdd{klozanov@illinois.edu}
\emailAdd{qianshu\_lu@g.harvard.edu}

\vspace*{1cm}

\abstract{In traditional models only an order one fraction of energy is transferred from the inflaton to radiation through nonperturbative resonance production in preheating immediately after inflation, due to backreaction effects. We propose a particle production mechanism that could improve the depletion of the inflaton energy density by up to four orders of magnitude. The improvement comes from the fast perturbative decays of resonantly produced daughter particles. They act as a ``spillway" to drain these daughter particles, reducing their backreaction on the inflaton and keeping the resonant production effective for a longer period. Thus we dub the scenario ``spillway preheating". We also show that the fraction of energy density remaining in the inflaton has a simple inverse power-law scaling in the scenario. In general, spillway preheating is a much more efficient energy dissipation mechanism, which may have other applications in model building for particle physics.}

\begin{document}

\maketitle

\section{Introduction}
\label{sec:intro}

Over the past decades, cosmological observations have provided compelling evidence for an inflationary phase in the early Universe and a hot big bang phase after it. Yet it is highly nontrivial to connect these two phases. It is generally believed that the phase transition is achieved through processes of (p)reheating, during which the inflaton energy is transferred to the thermal energies of other particles. The thermal particles could be produced either through perturbative decays of the inflaton \cite{Abbott:1982hn, Dolgov:1982th, Albrecht:1982mp}, or through non-perturbative and out-of-equilibrium dynamics \cite{Traschen:1990sw,Dolgov:1989us, Shtanov:1994ce, Kofman:1994rk, Boyanovsky:1995ud, Yoshimura:1995gc, Kaiser:1995fb, Kofman:1997yn,Allahverdi:2010xz,Amin:2014eta}. While the first possibility is called ``reheating", the latter possibility is often referred to as ``preheating", since it usually happens much faster and earlier than reheating.\footnote{At the end of preheating the daughter particles are not necessarily in thermal equilibrium (unlike in the case of reheating). However, they are in a `prethermal' state which has no memory of the initial conditions for preheating, set at the end of inflation \cite{Allahverdi:2010xz,Amin:2014eta,PhysRevLett.93.142002,Micha:2004bv}.}

Compared to reheating, preheating contains intriguing rich dynamics that is beyond the reach of perturbative calculations and calls for a better understanding. It could also lead to interesting direct or indirect observables such as a shift of inflation observables (scalar tilt and tensor-to-scalar ratio)~\cite{Liddle:2003as,Dai:2014jja,Munoz:2014eqa,Martin:2016oyk,Hardwick:2016whe,Lozanov:2016hid,Lozanov:2017hjm,Antusch:2020iyq}, a stochastic gravitational wave background at high frequencies~\cite{Khlebnikov:1997di,Easther:2006vd,Easther:2006gt,GarciaBellido:2007af,Dufaux:2007pt,Dufaux:2008dn,Dufaux:2010cf,Bethke:2013vca,Adshead:2018doq,Kitajima:2018zco,Bartolo:2016ami,Figueroa:2017vfa,Caprini:2018mtu,Bartolo:2018qqn,Lozanov:2019ylm,Adshead:2019igv,Adshead:2019lbr} as well as non-Gaussianities~\cite{Lyth:2001nq,Kofman:2003nx,Dvali:2003em,Chambers:2007se,Chambers:2008gu,Bond:2009xx,Leung:2012ve,Leung:2013rza,Imrith:2019njf,Fan:2020xgh}. Yet most preheating mechanisms that have been studied in the literature, such as parametric resonance~\cite{Kofman:1997yn} and tachyonic resonance~\cite{Dufaux:2006ee}, could at most transfer an order one fraction of the inflaton energy to radiation.\footnote{The only known exception to this rule of thumb is tachyonic gauge preheating with a scalar, $\varphi$, or a pseudo-scalar, $a$, inflaton coupled to the gauge field via $f(\varphi) F^2$ \cite{Deskins:2013dwa,Adshead:2017xll} or $aF\tilde{F}$ \cite{Adshead:2015pva,Cuissa:2018oiw} interaction terms, respectively. Such scenarios can boost the depletion of the inflaton energy density by up to two orders of magnitude.}\footnote{An analytical argument for the order one fraction of energy transfer, based on effective field theory, is given in Ref.~\cite{Giblin:2017qjp}.} In other words, they could not complete the phase transition from inflation to the thermal big bang. Perturbative reheating still needs to happen at a (much) later time to finish the transition. The central question, which is the focus of this paper, is then: {\it could there exist a new preheating mechanism to improve the efficiency of the energy transfer from the inflaton to radiation?}  

In this article, we propose a new preheating mechanism that could improve the depletion of the inflaton energy density by orders of magnitude, compared to the well-known mechanisms. The bottleneck of non-perturbative particle production is the backreaction effects. Once the (direct) daughter particles, e.g., scalars denoted as $\chi$'s, are copiously produced through various instabilities, they backreact on the inflaton, $\phi$, and pause the particle production processes. As a result, the inflaton releases at most about half of its energy to radiation, as realized in the tachyonic resonance preheating~\cite{Dufaux:2006ee}. 
One possible method to reduce the backreaction is to provide a {\it ``spillway"} to the daughter particles so that they could be drained after being produced abundantly and particle production could keep going without much backreaction. This could be realized through having the daughter particles decay perturbatively to second-generation daughter particles, e.g., fermions denoted as $\psi$'s. We will argue that the cascade decays, $\phi \to \chi \to \psi$ with the first step being non-perturbative and the second step being perturbative, could improve the depletion of the inflaton energy density by {\it up to four orders of magnitude}, within the range of parameters we could simulate numerically. We dub this new mechanism {\it spillway preheating}.

Alert readers may wonder why we cannot just have inflaton decay perturbatively, as in the simple reheating scenario, instead of combining the complicated particle production and the perturbative decays into a more complicated scenario? There are a couple of motivations to consider the spillway preheating: {\it a)} in this scenario, the perturbative decays $\chi \to \psi$ could happen on a much shorter time scale and thus during the preheating stage due to a larger coupling between $\chi$ and $\psi$, while the perturbative decays of the inflaton $\phi$ happens on a much longer time scale since the inflaton's couplings to other particles are usually suppressed (otherwise the inflaton's potential is unprotected during inflation\footnote{One could construct models where the inflaton's couplings vary during and after inflation \cite{Kofman:2003nx, Dvali:2003em, Bernardeau:2004zz}. We will not explore this possibility further.}); {\it b)} in a more realistic (p)reheating model containing the standard model (SM) beyond the simplified model we study involving only three species, one could easily envision such cascade decay processes: inflaton first produce some SM particles through preheating, e.g., $W$, $Z$ or $h$, which then subsequently decay into other SM particles.

Readers who are familiar with the preheating literature will note that spillway preheating has similar ingredients to a known scenario, instant preheating \cite{Felder:1998vq}. Yet our studies bear several important differences. In instant preheating, which we will review in more detail, perturbative decays of $\chi$'s occur during the first oscillation of the inflaton after inflation and end the preheating stage. It could again at most transfer an order one fraction of the inflaton energy to the daughter particles. Its main advantage is that it could be used to produce heavy particles which could not be generated through perturbative decays. As argued before, our goal is to improve the energy transfer efficiency of the preheating mechanism. We do {\it not} require the perturbative decays to happen during the first oscillation of the inflaton. Instead they kick in after ${\cal{O}}(1-10)$ oscillations when an order one fraction of energy in $\phi$ is transferred to $\chi$ and backreaction starts to become important. They convert $\chi$ into radiation that does not directly backreact on the inflaton. In this way, particle production without much backreaction could continue for a while until the driven instability disappears. As a net result, only a tiny fraction, as small as $10^{-4}$ of the total energy density remains in the inflaton while the dominant fraction is transferred to radiation.

The effects of perturbative decays had been studied in the framework of Higgs inflation in \cite{GarciaBellido:2008ab} and \cite{Repond:2016sol}. In those studies, the SM Higgs field is the inflaton and resonantly generates $W$ and $Z$ gauge bosons after inflation. The massive gauge bosons decay to SM fermions perturbatively. 
It was found that the fraction of Higgs inflaton energy density remained at the end of preheating reduces from $26\%$ to $2\%$ when the gauge boson decays are turned on. Our approach is similar to \cite{Repond:2016sol} in general, as will be explained later when we describe our model and numerical method. Yet instead of specifying the inflationary model and fixing relevant couplings as in \cite{Repond:2016sol}, we use a simplified model containing only the potential after inflation and allow the parameters to vary. It will be easier to derive and understand in the simplified model: {\it 1)} the constraints on the parameters for the spillway mechanism to work, and {\it 2)} the dependences of the results on the parameters. These results could be applied to particle production beyond the Higgs inflation scenario. In addition, we find that there exists regions of parameter space in which the remaining inflaton energy density can be reduced to as small as $0.01\%$.

The paper is organized as follows. In Sec.~\ref{sec:review}, we review the two preheating mechanisms in the literature which are most relevant to our scenario. In Sec.~\ref{sec:model}, we describe our model and the numerical approach to study the evolution of the system. In Sec.~\ref{sec:results}, we present our results. We demonstrate how the energy transfer efficiency is improved, study the parametric conditions for several key assumptions to hold and validate the results using different choices of parameters. We conclude in Sec.~\ref{sec:conc}.

\section{Non-perturbative particle production}
\label{sec:review}
In this section, we will review two of the non-perturbative particle production mechanisms, tachyonic resonance~\cite{Dufaux:2006ee} and instant preheating~\cite{Felder:1998vq}, which are most relevant to our study. We will summarize their key ingredients and features. Readers who are familiar with the subject could skip this section.

\subsection{Tachyonic resonance preheating}
\label{subsec:tach}
In the model of tachyonic resonance preheating~\cite{Dufaux:2006ee}, the potential after inflation is given by 
\beq
V_{\text{tach}} = \frac{1}{2}m^2\phi^2+\frac{1}{2}\frac{M^2}{f}\phi\chi^2+\frac{1}{4}\lambda\chi^4,
\label{eq:Vtach}
\eeq
where $\phi$ is the inflaton and $\chi$ is the scalar field that could be produced by $\phi$'s decays, either perturbatively or non-perturbatively. There are multiple mass and energy scales involved in the model. $m$ is the inflaton mass. Throughout the paper, we fix $m = 10^{-6} M_{\text{pl}}$ unless specified otherwise, with the reduced Planck scale $M_{\text{pl}} \approx 2.4 \times 10^{18}$ GeV, which is chosen to be consistent with the CMB constraints on inflation~\cite{Akrami:2018odb}. The high energy scale $f$ that suppresses the coupling between $\phi$ and $\chi$ will be taken to be the Planck scale, $f= M_{\text{pl}}$. After inflation, $\phi$ starts to oscillate around the minimum of its potential with an oscillation amplitude $\Phi$. The initial amplitude is taken to be $\Phi_0 =f$. As $\phi$ oscillates, $\chi$ obtains an effective mass through its coupling to $\phi$, which is of order $M$ initially. When embedding this toy model in a UV completion, e.g., a supersymmetric scenario, $M$ and $m$ both originate from SUSY breaking and are of the same order without tuning in the simplest case. Yet for the particle production to happen, we need $M/m \gtrsim {\cal{O}}(10)$, which we will explain below. This requires either tuning in the UV theory or a more complicated model, e.g., a model in which SUSY breaking contributing to $m$ is sequestered compared to that to $M$. $\lambda$ is the self-coupling of $\chi$ and the self-coupling term is needed for the potential to be bounded from below.

\begin{figure}[h]
    \centering
\includegraphics[width=0.5\textwidth]{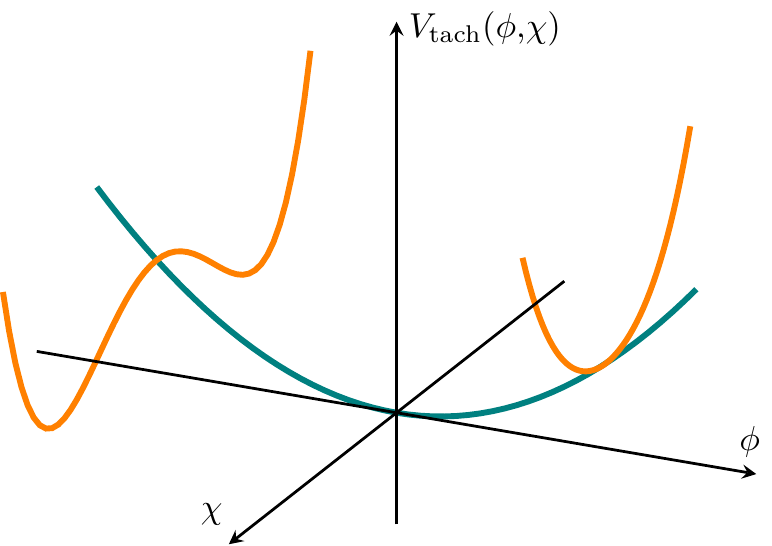}
    \caption{The shape of the potential $V_{\text{tach}}$ in Eq.~\eqref{eq:Vtach}.}
    \label{fig:potential}
\end{figure}

The potential is sketched in Fig.~\ref{fig:potential}. As $\phi$ oscillates around the minimum of its potential, the effective mass squared of $\chi$, $M^2\phi/f$, also oscillates. When $\phi$ passes through the origin from the positive to the negative side, the sign of the mass squared term flips and triggers a tachyonic instability, which could drive an exponentially fast production of $\chi$. At the initial stage of particle production when only a small fraction of energy is transferred from $\phi$ to $\chi$, one could study the system using linearized equations of motion and, e.g., carry out a Floquet analysis. Such stability analysis show that for the tachyonic instability to develop, we need
\beq
q_0 = \frac{M^2}{m^2} \gg 1. 
\label{eq:q0}
\eeq
A more intuitive way to understand this requirement is by comparing two different time scales. The oscillation period of the inflaton is $\sim 1/m$ while the time scale associated with the change of $\chi$'s potential is $\sim 1/M$. In order for $\chi$ to respond to the change of its potential within one oscillation of the inflaton, i.e., to be excited non-adiabaticlly, we need $1/m \gg 1/M$, which is equivalent to Eq.~\eqref{eq:q0}. 

Once there are comparable energies in $\phi$ and $\chi$, the backreaction from $\chi$ to $\phi$ could no longer be ignored and the linear approximation breaks down. One way to see that is when $\langle \chi^2 \rangle$ is sufficiently large, the self-coupling term of $\chi$, that is ignored in the linear analysis, turns into a positive effective mass term: $\lambda \langle \chi^2 \rangle \chi^2$, which becomes increasingly important and always counteracts the trilinear coupling when $\phi$ flips to the tachyonic side. The larger $\lambda$ is, the stronger the backreaction becomes and the less efficient particle production is. 
The detailed evolution of the system has to be studied by numerical simulations. Yet the final energy transfer efficiency is roughly controlled by a single parameter, the backreaction efficiency parameter, 
\beq
b \equiv \frac{1}{4} \left(\frac{\frac{1}{2} \frac{M^2}{f} \phi \chi^2}{\frac{1}{2} m^2 \phi^2}\right) \left(\frac{\frac{1}{2} \frac{M^2}{f} \phi \chi^2}{\frac{1}{4} \lambda \chi^4}\right)= \frac{M^4}{2\lambda m^2 f^2}. 
\label{eq:b}
\eeq
We need $b \leq 1$ for the potential to be bounded from below (so that $\lambda$ could not be zero or arbitrarily small), and $b\approx 1$ for tachyonic resonance to be efficient. 

When $q_0 \gg 1$ and $b \approx 1$, the maximum fraction of energy transferred from $\phi$ to $\chi$ is about $50\%$. The equation-of-state of the coupled system reaches a plateau with $w \subset (0.2 - 0.3)$, signaling an exotic and intriguing mixed matter-radiation epoch. 
For more detailed discussions on the parametric conditions in Eq.~\eqref{eq:q0} and \eqref{eq:b} as well as numerical analyses, see Refs.~\cite{Amin:2018kkg} and~\cite{Fan:2020xgh}.

\subsection{Instant preheating}
\label{subsec:inst}
In instant preheating~\cite{Felder:1998vq}, particle production is achieved almost instantaneously within the first oscillation of the inflaton. In the original model, the potential is given by 
\beq
V_{\rm{ins}} = \frac{1}{2}m^2\phi^2+\frac{g^2}{2} \phi^2 \chi^2 +y\chi\bar{\psi}\psi,
\label{eq:Vins}
\eeq
where $\psi$ is a heavy fermion field and $y$ is the Yukawa coupling. 

When $\phi$ crosses the origin in the field space in the first oscillation, particle production of $\chi$ occurs, the same as what happens in the well-known parametric resonance preheating~\cite{Kofman:1997yn}. The effective mass of $\chi$, proportional to $\phi^2$, is initially small when being produced and keeps growing as $\left|\phi\right|$ increases. Thus the energy density stored in $\chi$ grows as well. When $\left|\phi\right|$ gets close to the maximum value (the initial oscillation amplitude), $\chi$ decays to $\psi$'s and dumps all its energy into the fermions. Note that in order to achieve this mechanism, one must finely tune $y$ so that the perturbative decay lifetime of $\chi$ matches the (quarter of the) oscillation period of $\phi$~\cite{Felder:1998vq}:
\beq
y^2 g \approx 5 \times 10^{-4}. 
\eeq

We want to emphasize that this simplest model could still at most transfer an order-one fraction of energy from the inflaton to the daughter fields, which is only achieved when the quartic coupling $g$ is of order one. If this order-one quartic coupling is present during inflation, it will spoil the flatness of the inflaton potential in the absence of fine tuning. To avoid tuning the inflaton potential, one then needs to construct models in which the coupling is absent during inflation and is only effective after inflation. In short, from the point of view on the efficiency of energy transfer, instant preheating does {\it not} improve over other preheating mechanisms, i.e., tachyonic resonance preheating. Its main advantage is that it could produce very heavy particles (e.g., heavy $\psi$'s in Eq.~\eqref{eq:Vins}) with masses which could be close to the Planck scale.

\section{Model and approach}
\label{sec:model}
In this section, we will present our model and describe the approach to simulate the evolution of the system. In our model, the potential after inflation is given by 
\beq
V_{\rm{spillway}} = \frac{1}{2}m^2\phi^2+\frac{1}{2}\frac{M^2}{f}\phi\chi^2+\frac{1}{4}\lambda\chi^4+y\chi\bar{\psi}\psi,
\eeq
where the notations are the same as in the previous section: $\phi$ is the inflaton; $\chi$ is a scalar while $\psi$ is a fermion. 

At first glance, the model is simply a hybrid of the tachyonic resonance and instant preheating. The motivation to consider this model is as follows. While tachyonic resonance is very efficient (more efficient than parametric resonance) in energy transfer, it could still at most transfer about half of the inflaton energy to the daughter fields due to backreaction once a large number of $\chi$'s are produced. The reason for the addition of the perturbative decays $\chi \to \bar{\psi} \psi$ is to drain $\chi$'s to reduce the backreaction from $\chi$ to $\phi$ and to keep the tachyonic production going. It is then expected that the cascade decays $\phi \to \chi \to \psi$ could improve the energy transfer from the inflaton to radiation. 
Indeed as we will show in the next section, the evolution of the system behaves quite differently, with significantly improved energy transfer efficiency in at least part of the parameter space of this model, compared to the two models reviewed in the previous section.

The Yukawa coupling generates a nonzero $\chi$ decay width, which we approximate as
\beq
\Gamma_{\chi} = \frac{y^2}{8\pi}m_{\chi}(\phi),
\label{eq:width}
\eeq
where the effective mass of $\chi$, $m_{\chi}$, is $\phi$-dependent due to the trilinear coupling between $\phi$ and $\chi$. To ensure that $m_{\chi}$ is real and positive at all times, we will define $m_\chi$ to be the curvature at the minimum of its potential, which is
\beq
\Gamma_{\chi} = \begin{cases}
\frac{y^2}{8\pi} \sqrt{\frac{M^2}{f}\phi}, &\phi > 0\\
\frac{y^2}{8\pi} \sqrt{\frac{2M^2}{f}\abs{\phi}}, &\phi < 0.
\end{cases}
\eeq
For the same $|\phi|$, $m_{\chi}$ is larger by a factor of $\sqrt{2}$ for $\phi<0$ because $V_{\text{tach}}$ has a higher curvature at the minimum of the double-valley than the single-valley as shown in \Fig{fig:potential}.

Conventionally, numerical study of the preheating system is implemented by solving the equations of motion for the classical fields and Friedmann equations on a spatial lattice. The classical field equations are good approximations when the occupation numbers of the fields are high. But our system contains a fermion field, whose evolution may not be well approximated by its classical equation of motion. Moreover, the classical field equation for $\chi$ with the potential above will not generate the correct behavior for a field with a nonzero perturbative decay width. For a decaying field, the occupation number with momentum $k$ should decrease exponentially as
\beq
n_{\chi}(k)\sim \exp[-\frac{\Gamma_{\chi} t}{\gamma}],
\eeq
where $\gamma$ is the boost factor associated with the momentum $k$. In particular, the rate of the exponential decay is dependent only on $m_{\chi}(\phi)$, while in the classical equation of motion for $\chi$, the Yukawa term will contribute a $\psi$-dependent term instead.

To overcome the difficulties above, we adopt the strategy used in \cite{Repond:2016sol}. The equation of motion for the inflaton, $\phi$, is not affected by the Yukawa coupling and its induced decays: 
\beq \ddot{\phi}+3H\dot{\phi}-\frac{1}{a^2}\nabla^2\phi+m^2\phi+\frac{M^2}{f}\chi^2 = 0,
\label{eq:eomphi}
\eeq
where $a$ is the scale factor. 
Then, to mimic the effect of $\chi$ decays, we add a friction term, $\Gamma_{\chi}\dot{\chi}$, to the equation of motion for $\chi$: 
\beq \ddot{\chi}+3H\dot{\chi}-\frac{1}{a^2}\nabla^2\chi+\frac{M^2}{f}\phi\chi+\lambda\chi^3+\Gamma_{\chi}\dot{\chi} = 0,
\label{eq:eomchi}
    \eeq
where $\Gamma_{\chi}$ is given in Eq.~\eqref{eq:width}. The fermionic decay products are modeled as a homogenous, radiation-like perfect fluid with energy density $\rho_{\psi}$, which is independent of position. The time evolution equation for $\rho_{\psi}$ is derived by the conservation of the stress-energy tensor of the entire system including $\phi$, $\chi$ and the fermionic fluid, combined with Eqs.~\eqref{eq:eomphi} and \eqref{eq:eomchi}: 
    \beq
    \nabla_{\mu}T^{\mu0} = 0\quad \Rightarrow \quad \dot{\rho}_{\psi}+4H\rho_{\psi} - \langle \Gamma_{\chi} \dot{\chi}^2 \rangle = 0,
   \label{eq:fluid}
    \eeq
 where in the last term, $\langle \cdots \rangle$ refers to the spatial average. 
We see that the $\dot{\chi}^2$ term will act as a perpetual source of $\rho_{\psi}$, which is what we expect since $\chi$ decays to $\psi$'s. But because $\chi$ is not a homogeneous field, forcing its decay products to be a homogenous fluid means that the conservation equations $\nabla_{\mu}T^{\mu i} = 0$ are violated. In other words, the gradient energy of $\chi$ is lost when it decays to $\rho_{\psi}$. Whether this is a large effect or not can be checked with self-consistency in the background evolution of the Universe. Evolution of the scale factor is governed by
\beq
\frac{\ddot{a}}{a} = -\frac{4 \pi G}{3}\expval{\rho_{\text{tot}}+3p_{\text{tot}}},\quad \left(\frac{\dot{a}}{a}\right)^2 = \frac{8\pi G}{3}\expval{\rho_{\text{tot}}},
\label{eq:scale}
\eeq
where $\rho_{\text{tot}}$ and $p_{\text{tot}}$ are the total energy density and pressure density, including contribution from the fermion fluid. These two equations are redundant when combined with the two equations for $\phi$ and $\chi$ as well as the evolution equation of the fermionic fluid. We will use the second scale factor equation as a consistency check for ``energy conservation''. In all simulations, energy conservation is satisfied at the level of $10^{-3}$ or higher precision. This means that the approximation of $\rho_{\psi}$ as a homogenous fluid does not generate a large loss of gradient energy.

To solve Eqs.~\eqref{eq:eomphi}, \eqref{eq:eomchi}, \eqref{eq:fluid} and \eqref{eq:scale}, we use the public LatticeEasy package~\cite{Felder:2000hq}, but with the integrator modified to the 4th order Runge-Kutta method. 

If the perturbative decay products of $\chi$ were scalars instead of fermions, we would have a different phenomenology. For example, if a scalar field, $\varphi$, is coupled to $\chi$ through the trilinear interaction $V_{\rm int}=\sigma\chi\varphi^2$, where $\sigma$ is a dimensionful coupling constant, the perturbative decay rate is $\Gamma_{\chi\rightarrow\varphi\varphi}=\sigma^2/(8\pi m_{\chi}(\phi))$ \cite{Kofman:1997yn,Dufaux:2006ee}. Note that the dependence of the decay rate on the $\chi$ effective mass, $\propto m_\chi^{-1}$, is opposite to the one in the case of a fermionic daughter field, see Eq. \eqref{eq:width}. Hence, $\chi$ would be most likely to decay when it has a vanishing mass, which coincides with the zero-crossings of $\phi$. At these moments, $\chi$ evolves non-adiabatically. Its time-dependent ground and excited states do not coincide with the ones of the free theory in flat spacetime and thus its decays into pairs of $\varphi$'s cannot be captured with the phenomenological friction term. On the other hand, in the fermion case the perturbative decays of $\chi$ occur in the adiabatic regime (when $\phi$ is near the extreme or evolves slowly) and can be described as the exponential damping of the amplitude of the excited $\chi$ (whose ground and excited states are the ones of the free theory in flat spacetime). Nevertheless, the non-perturbative decays of $\chi$ into $\varphi$ pairs can be studied with classical lattice simulations with all the scalar fields included \cite{Dufaux:2006ee}. However, our numerical simulations show that this scenario does not lead to improved efficient energy transfer to the daughter fields. 

Alternatively, we can consider a scalar coupling of the form $V_{\text{int}} \supset y\chi\varphi^3$. The decay width of $\chi$ will then be proportional to its mass, $\Gamma_{\chi\to \varphi\varphi} \propto y^2m_{\chi}(\phi)$, similar to the decays through the Yukawa coupling in our model. But compared to the $\phi$-$\chi$-$\psi$ system, the stability condition of the three-scalar system is more complicated, because both $\chi$ and $\varphi$ directions are potentially unstable depending on the sizes of their individual quartic self-interaction strengths. The stability condition can no longer be defined using a single parameter like we do in \Eq{eq:b}. In order to reduce the number of moving parts, we only consider fermionic perturbative decay products of $\chi$ in this paper.

\section{Results}
\label{sec:results}
We will present and discuss the results of the numerical simulations for our model in this section. We will first show the key results in one class of benchmark parameters and compare them with those of tachyonic resonances without perturbative decays. We will then discuss one key assumption of our simulations, that is, ignoring the backreaction of the fermionic fluid back to the scalar sub-system. Based on analytical arguments, we provide parametric relations between different parameters of the model for the assumption to hold. Lastly, we revisit our benchmark parameter choices and discuss the alternative choice as well as the validity of the results.

\subsection{Enhanced energy transfer}
\label{subsec:enhanced}
As reviewed in Sec.~\ref{subsec:tach}, to have efficient particle production in the tachyonic resonance scenario, we want to satisfy $q_0 \gg 1$ and $b \sim 1$. Motivated by these relations, we first choose
\beq
q_0 = \frac{M^2}{m^2} = 200, \quad b =\frac{M^4}{2\lambda m^2 f^2}= 0.9.  
\eeq
We also take $m = 10^{-6} \MP$ and the initial amplitude of the inflaton $\Phi_0 = f = \MP$, as explained before. Note that this set of parameters corresponds to a tiny quartic coupling $\lambda \approx 3.3 \times 10^{-8}$.

\begin{figure}[h]
    \begin{subfigure}[b]{0.49\textwidth}
    \centering
    \includegraphics[width=\textwidth]{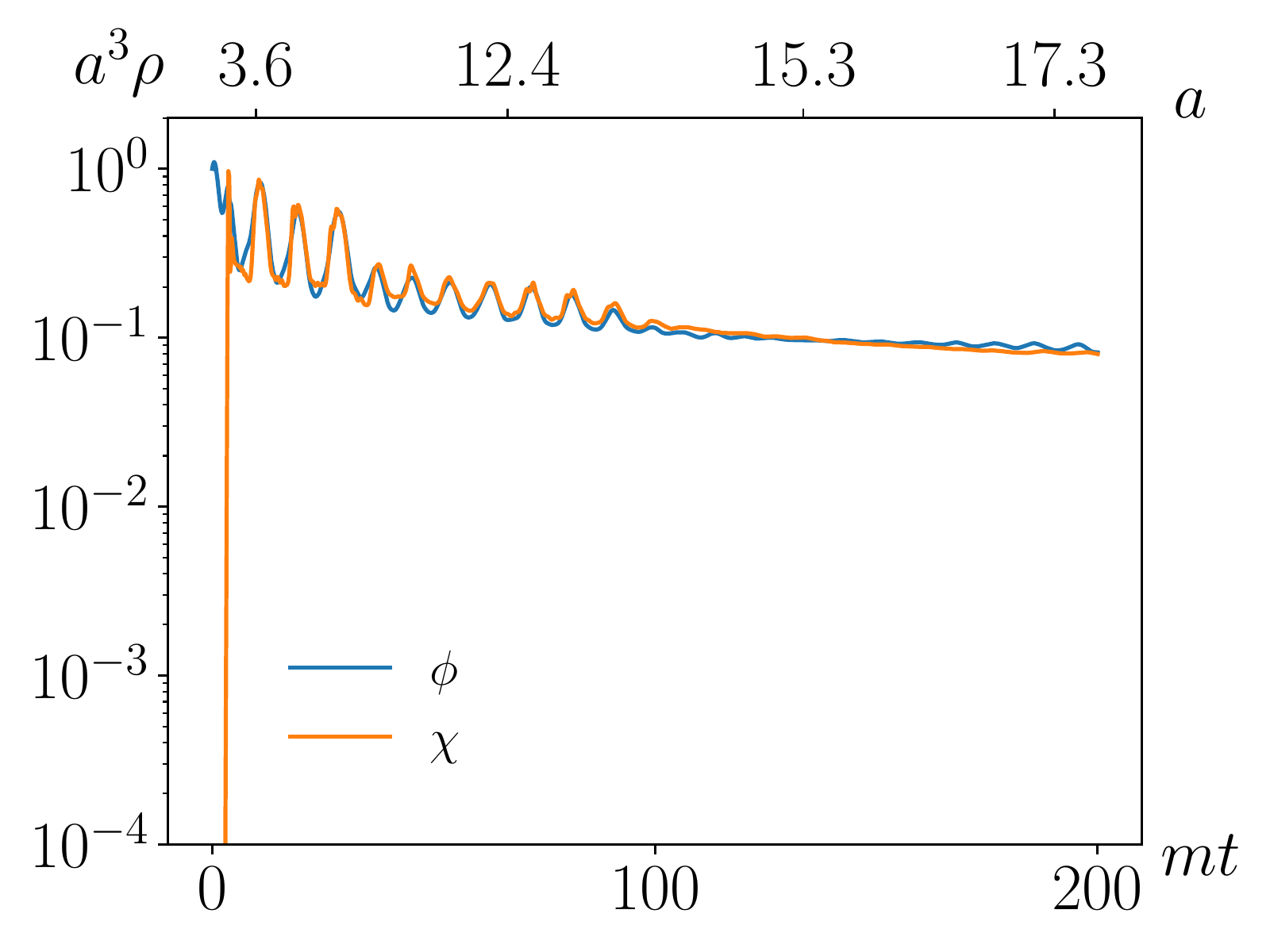}
    \caption{$y^2/8 \pi = 0$}\label{fig:q200_0_rhoEvol}
    \end{subfigure}
    \hfill
    \centering
    \begin{subfigure}[b]{0.49\textwidth}
    \centering
    \includegraphics[width=\textwidth]{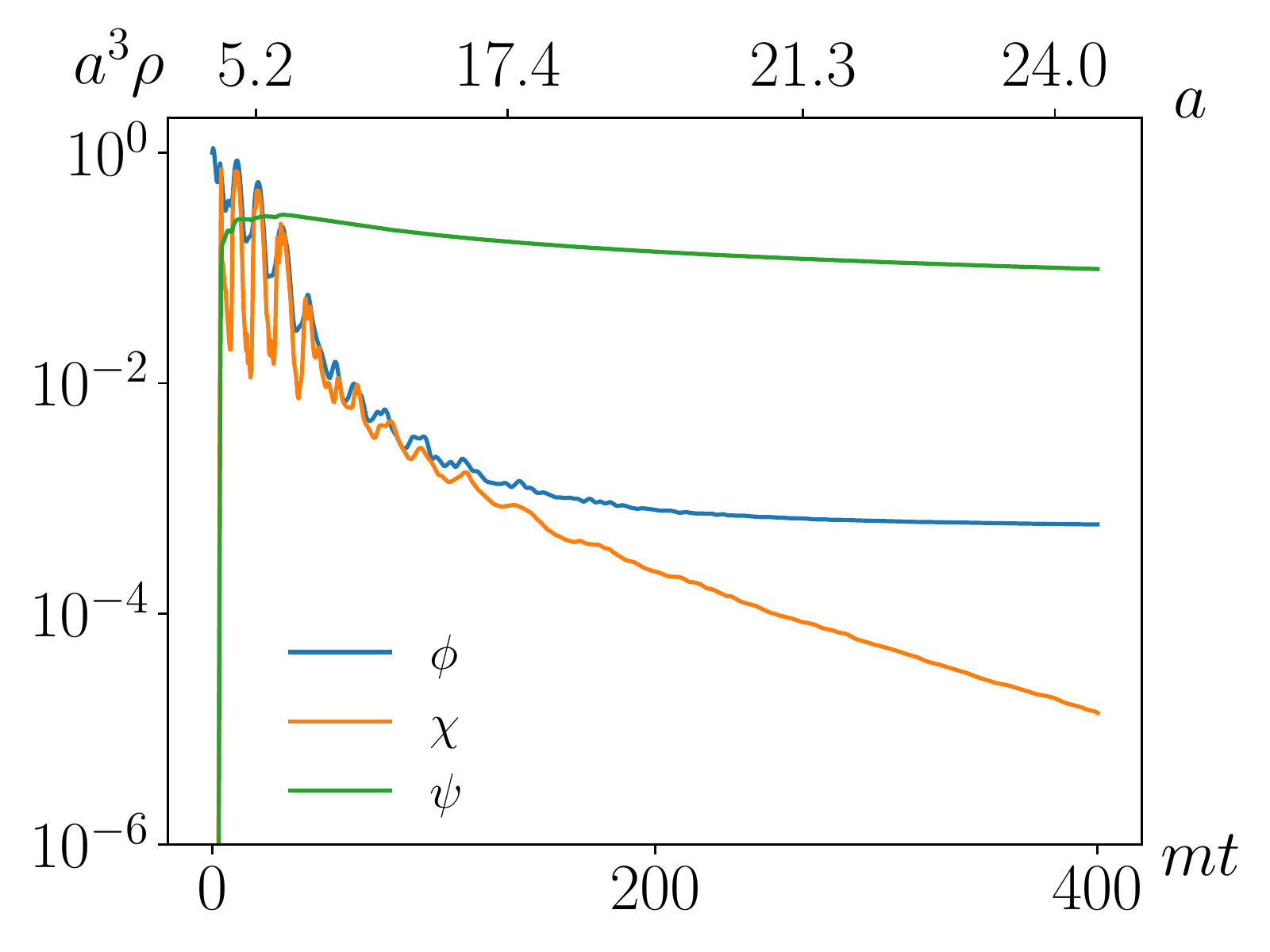}
    \caption{$y^2/8 \pi = 0.1$}\label{fig:q200_010_rhoEvol}
    \end{subfigure} 
    \caption{Time evolution of $\phi$, $\chi$, and fermionic fluid energy density for $b = 0.9$, $m = 10^{-6} \MP$, $\Phi_0 = f = \MP$, $q_0 = 200$ and $y^2/8\pi = 0$ or 0.1. When $y$ is nonzero, energy transfer continues to happen after $\rho_{\chi}$ becomes comparable with $\rho_{\phi}$. This second stage of energy transfer stops eventually when the system leaves the tachyonic resonance band. The smaller $y$ is, the longer it takes to reach this stop. The total comoving energy density is not conserved here because the equation of state of the system quickly deviates from being matter-like, as shown in \Fig{fig:small_lam_w}.}\label{fig:small_lam_rhoEvol}
\end{figure}

\begin{figure}[h]
    \begin{subfigure}[b]{0.49\textwidth}
    \centering
    \includegraphics[width=\textwidth]{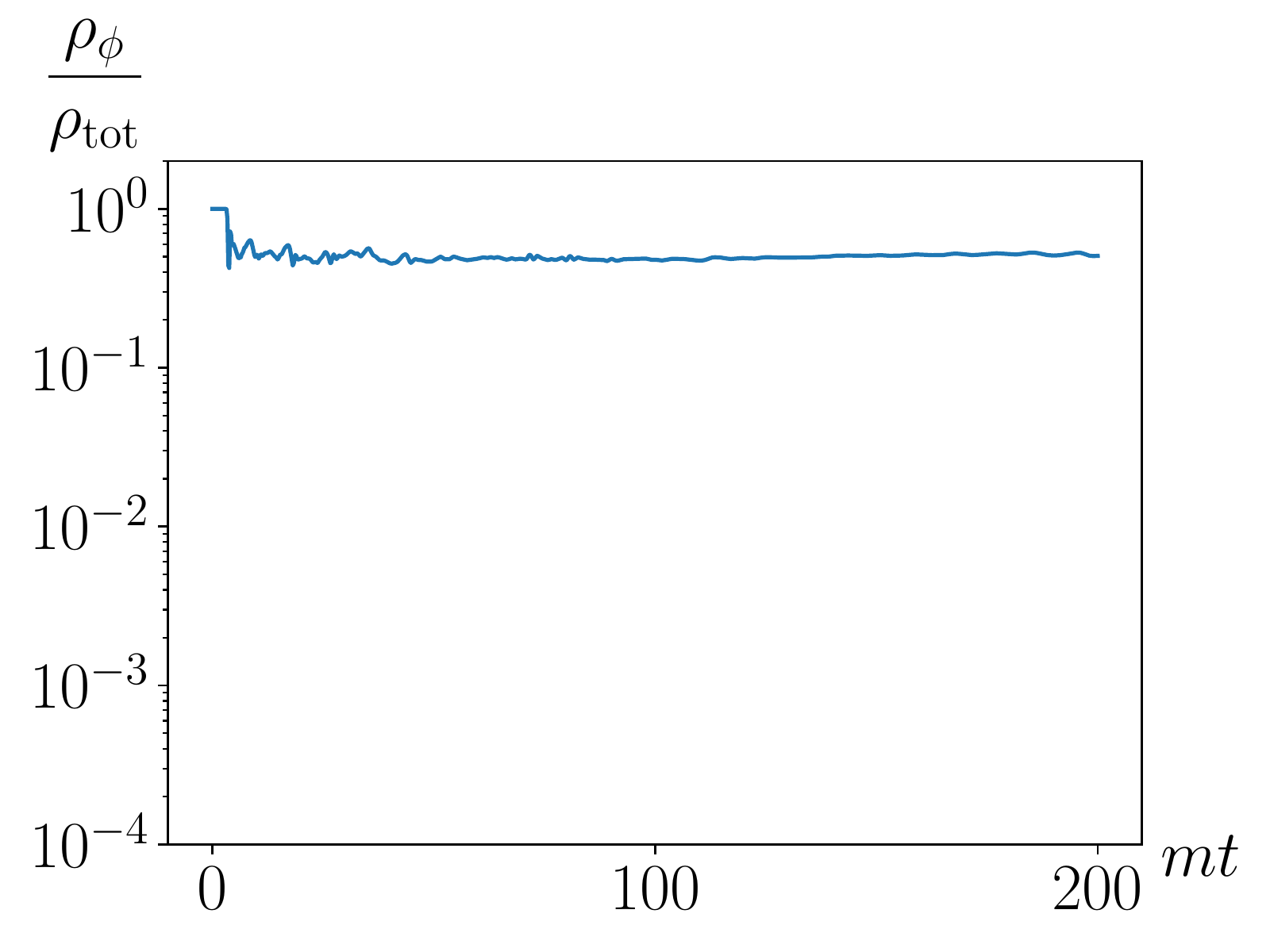}
    \caption{$y^2/8 \pi = 0$}\label{fig:q200_0_rhoRatio}
    \end{subfigure}
    \hfill
    \begin{subfigure}[b]{0.49\textwidth}
    \centering
 \includegraphics[width=\textwidth]{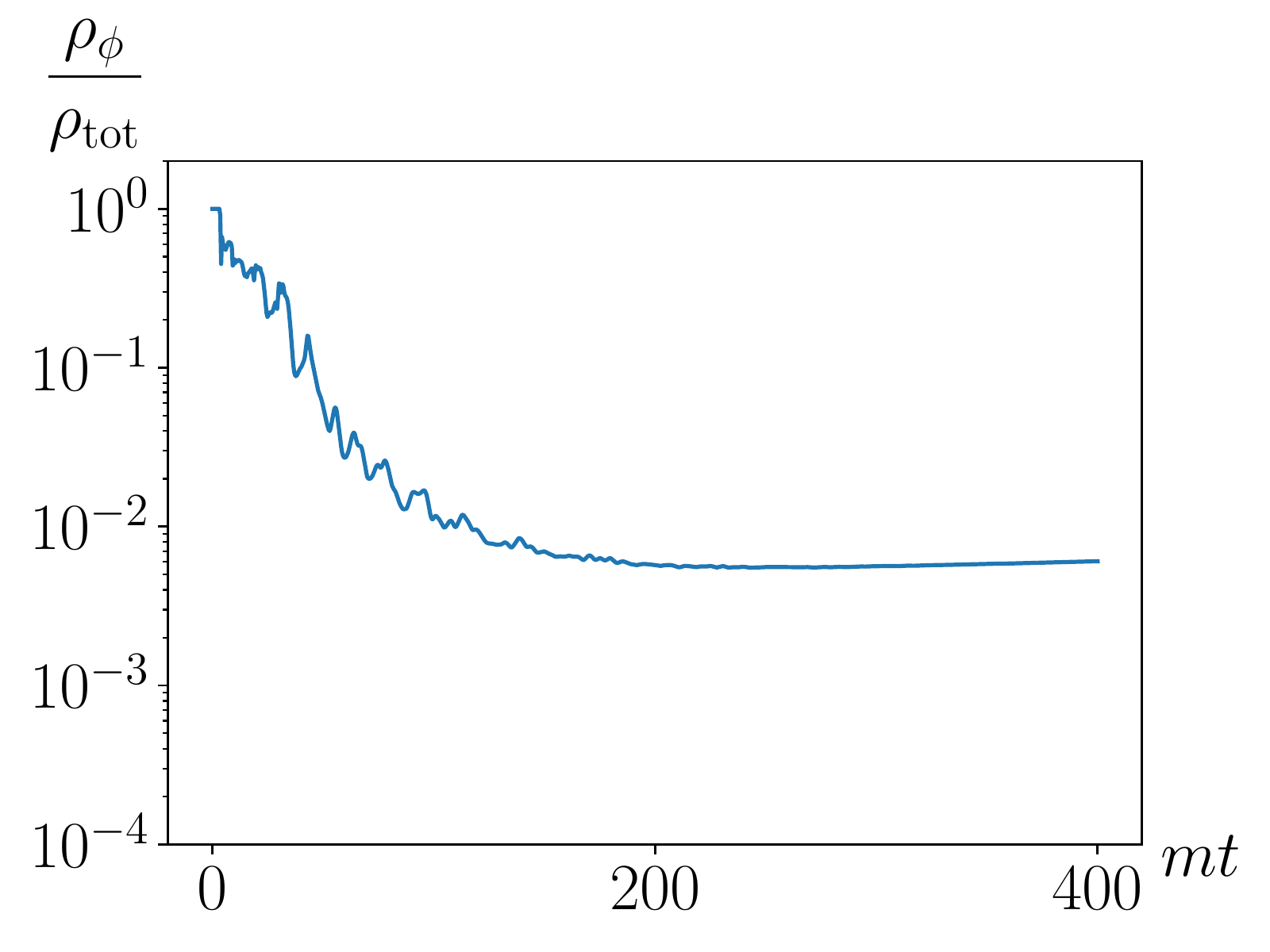}
    \caption{$y^2/8 \pi = 0.1$}\label{fig:q200_010_rhoRatio}
    \end{subfigure}
    \caption{Time evolution of $\rho_{\phi}/\rho_{\text{tot}}$ for $b = 0.9$, $m = 10^{-6} \MP$, $\Phi_0 = f = \MP$, $q_0 = 200$ and $y^2/8\pi = 0$ or 0.1. The initial rapid decrease is due to tachyonic production of $\chi$. When $y \neq 0$, the $\chi\to \bar{\psi}\psi$ decay alleviates $\chi$'s backreaction on $\phi$, and $\rho_{\phi}/\rho_{\text{tot}}$ continues to decrease. After the second stage of energy transfer ends, $\rho_{\phi}/\rho_{\text{tot}}$ slowly increases since $\phi$ is matter-like and redshifts slower than the rest of the system, which is radiation-like.}\label{fig:small_lam_rhoRatio}
\end{figure}

We simulate the system on a box of length $L =2 m^{-1}$ with $128^3$ points and $y^2/(8\pi) = 0$ and 0.1. We also put a UV cutoff on the initial power spectra of $\phi$ and $\chi$ to keep only modes that are excited by the linear tachyonic resonance. In other words, we cut off the initial spectrum of $\phi$ at $k_{\phi,\text{max}}/m = 0$ and we cut off the spectrum of $\chi$ at $k_{\chi,\text{max}}/m = 2\sqrt{q_0}$. The time evolutions of energy densities of $\phi$, $\chi$, and fermionic fluid, in the comoving volume, are shown in Fig.~\ref{fig:small_lam_rhoEvol}. We use $\rho_{\phi}/\rho_{\text{tot}}$ as a measure of the efficiency of energy transfer, where $\rho_{\text{tot}}$ is the total energy density of the system. The time evolution of $\rho_{\phi}/\rho_{\text{tot}}$ is shown in Fig.~\ref{fig:small_lam_rhoRatio}.

From the simulation results, we see that the $y = 0$ and $y\neq 0$ systems exhibit qualitatively different features in the time evolution of energy densities. To understand the differences, let's first check what happens when $y = 0$. In the beginning, $\rho_{\chi}\ll \rho_{\phi}$, and the system is in the linear regime. Energy rapidly transfers from $\phi$ to $\chi$ by tachyonic resonance production, and $\rho_{\chi}$ becomes comparable with $\rho_{\phi}$ within $mt\sim \mathcal{O}(1)$. Then the two fields evolve nonlinearly for a long time while maintaining $\rho_{\chi}\approx \rho_{\phi}$. Thus $\rho_{\phi}/\rho_{\text{tot}}$ stays around $\approx 0.5$, as shown in Fig.~\ref{fig:q200_0_rhoRatio}. The fact that $\rho_{\phi}/\rho_{\text{tot}}$ stays relatively constant after the system enters the non-linear regime does not mean that energy transfer from the inflaton stops. If the energy transfer $\phi\to \chi$ is completely shut off, $\rho_{\phi}/\rho_{\text{tot}}$ would increase since $\rho_{\phi}$ is matter-like and redshifts slower than $\rho_{\chi}$, which is radiation-like. As soon as the expansion of the Universe reduces the ratio $\rho_{\chi}/\rho_{\phi}$, tachyonic resonance quickly transfers energy to increase the ratio to reach $\rho_{\chi}\approx \rho_{\phi}$ again. In other words, when $\rho_{\chi}$ becomes comparable with $\rho_{\phi}$, the backreaction of $\chi$ does not terminate the tachyonic resonance, it simply {\it pauses} it. 
Making this distinction is important to understand the system with $y\neq 0$. The system with the cascade decays $\phi \to \chi \to \psi$, is precisely exploiting the fact that rapid $\phi\to\chi$ transfer will happen again once we reduce the energy density of $\chi$. This is evident from the evolution of energy densities shown in Fig.~\ref{fig:q200_010_rhoEvol}. After the initial rapid growth of $\rho_{\chi}$ to reach $\rho_{\chi}\approx \rho_{\phi}$, the $\chi\to \bar{\psi}\psi$ decays continuously reduce $\rho_{\chi}$, alleviate backreaction of $\chi$, and restart $\phi\to\chi$ until the energy equipartition between $\phi$ and $\chi$ is achieved again. Indeed we see that $\rho_{\phi}$ closely tracks $\rho_{\chi}$ for a while before decoupling. Thanks to the continuous draining of $\rho_\chi$ from $\chi \to \bar{\psi}\psi$ decay, the energy transfer from $\phi$ to $\chi$ is still rapid during the nonlinear regime, and this helps achieve the depletion of the inflaton energy density improved by two orders of magnitude compared to the $y = 0$ case.

For both cases, the tachyonic resonance will eventually be terminated when the value of $\phi$ is too small to drive the resonance. The value of $\phi$ could be reduced by both the redshift due to the expansion of the Universe and decays to the daughter fields. When there is no longer resonant production, no more energy will be transferred out of $\phi$, and the time evolution of $\rho_{\phi}$ decouples from that of $\rho_{\chi}$. The ratio $\rho_{\phi}/\rho_{\text{tot}}$ reaches the minimum value at this point. After $\phi$ decouples, $\rho_{\phi}$ redshifts slower than the rest of the system and $\rho_{\phi}/\rho_{\text{tot}}$ gradually increases from the minimum value. This general picture of what happens after tachyonic resonance terminates apply to both systems with $y = 0$ or $y\neq 0$. But when the termination happens can be drastically different. For system with $y = 0$, once the backreaction is effective, $\phi$ and $\chi$ are always coupled together, and there is no significant decrease in $a^3 \rho_\phi$. The decoupling may not happen until well beyond the time range we could simulate. However, for $y\neq 0$, the rapid decrease in $a^3\rho_{\phi}$ implies a quick reduction in $a^{3/2}\phi$, and tachyonic resonance can potentially end much earlier. Indeed as shown in Fig.~\ref{fig:q200_010_rhoEvol}, for $y^2/(8\pi) = 0.1$, $\phi$ decouples from $\chi$ at around $mt\approx 100$. This is also confirmed in Fig.~\ref{fig:q200_010_rhoRatio}: $\rho_{\phi}/\rho_{\text{tot}}$ decreases rapidly initially and then reaches a minimum value at around  $mt\approx 100$. After decoupling, it increases slowly due to the redshift effects. 

When $y \neq 0$, even though redshifts can slowly increase the fraction of inflaton energy back up again, it is important to understand how the minimum value of $\rho_\phi/\rho_{\rm tot}$ achieved scales with different parameters in the model. A full analytic understanding of the relationship is difficult given the nonlinearity of the evolution. Yet we could still learn something useful from the linear analysis. In the Floquet analysis, the tachyonic instability exists only when
\beq
q = \frac{M^2}{m^2}\frac{\Phi}{f} = q_0\frac{\Phi}{f} \gg 1,
\eeq
where $\Phi$ is the coherent oscillation amplitude of $\phi$. 
For a given $q_0$, tachyonic resonance terminates when $\Phi/f\sim 1/q_0$. The fraction of inflaton energy density at this point is
\beq
\frac{\rho_{\phi}}{\rho_{\text{tot}}} \sim \frac{\Phi^2}{f^2}\sim \frac{1}{q_0^2}.
\eeq
This linear analysis is not going to be a precise description of the evolution, but we expect the conclusion holds generally: when $q_0$ is larger, more energy is transferred from $\phi$, and a smaller $(\rho_{\phi}/\rho_{\text{tot}})_{\text{min}}$ value is achieved.

\begin{figure}[th]
    \centering
    \begin{subfigure}[b]{0.45\textwidth}
         \centering
         \includegraphics[width=\textwidth]{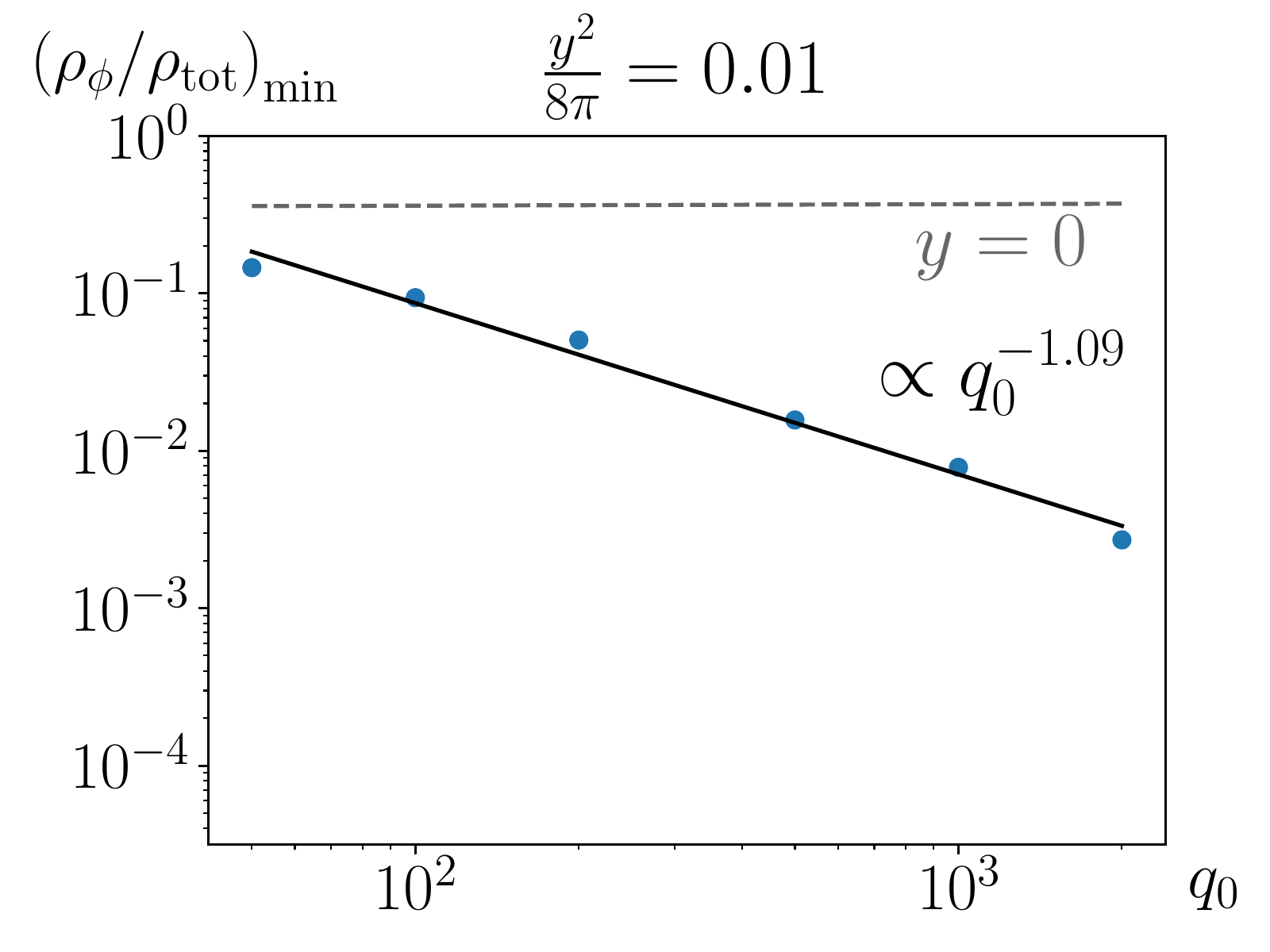}
     \end{subfigure}
     \begin{subfigure}[b]{0.45\textwidth}
          \centering
          \includegraphics[width=\textwidth]{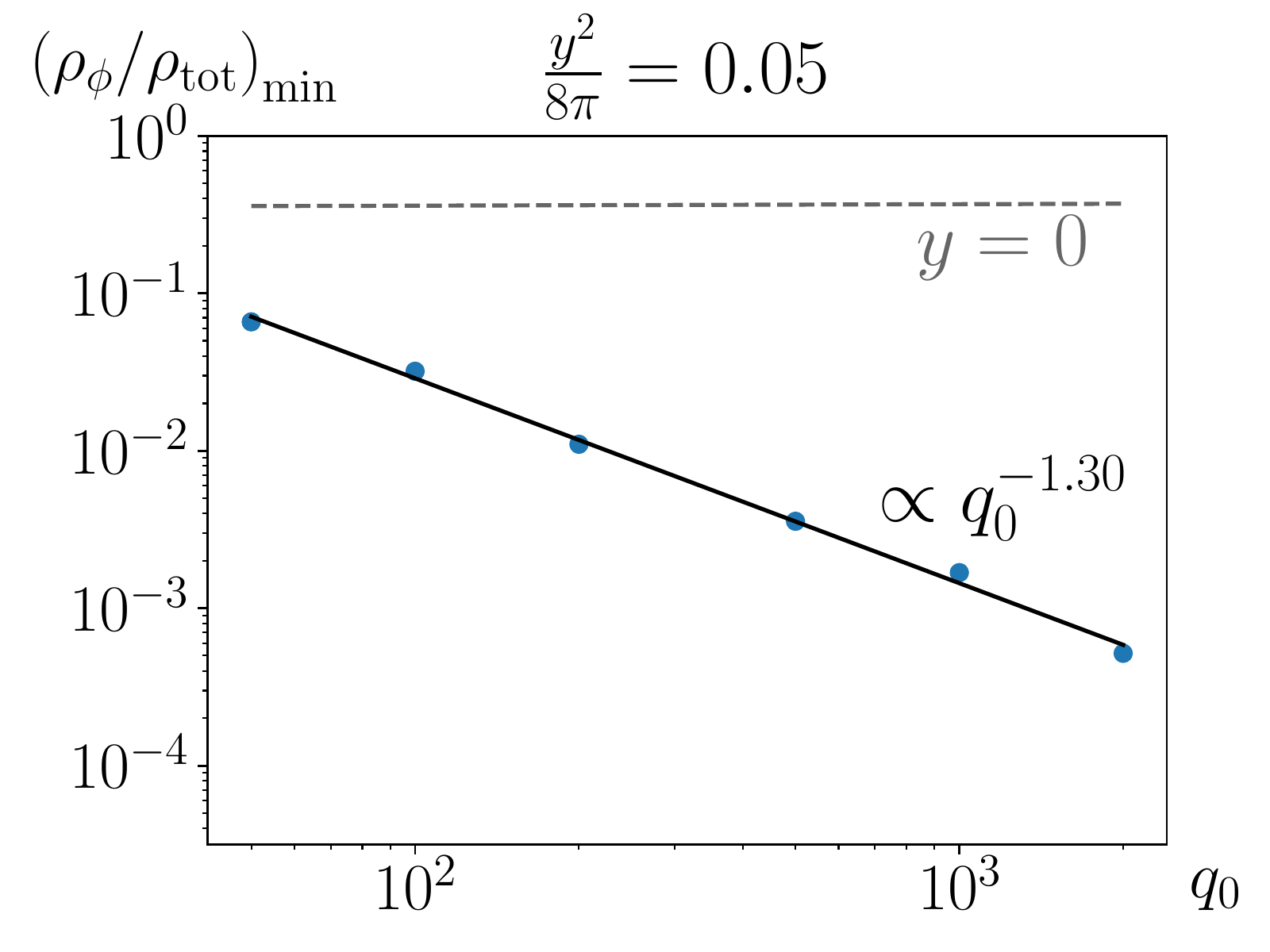}
      \end{subfigure}
      \begin{subfigure}[b]{0.45\textwidth}
           \centering
           \includegraphics[width=\textwidth]{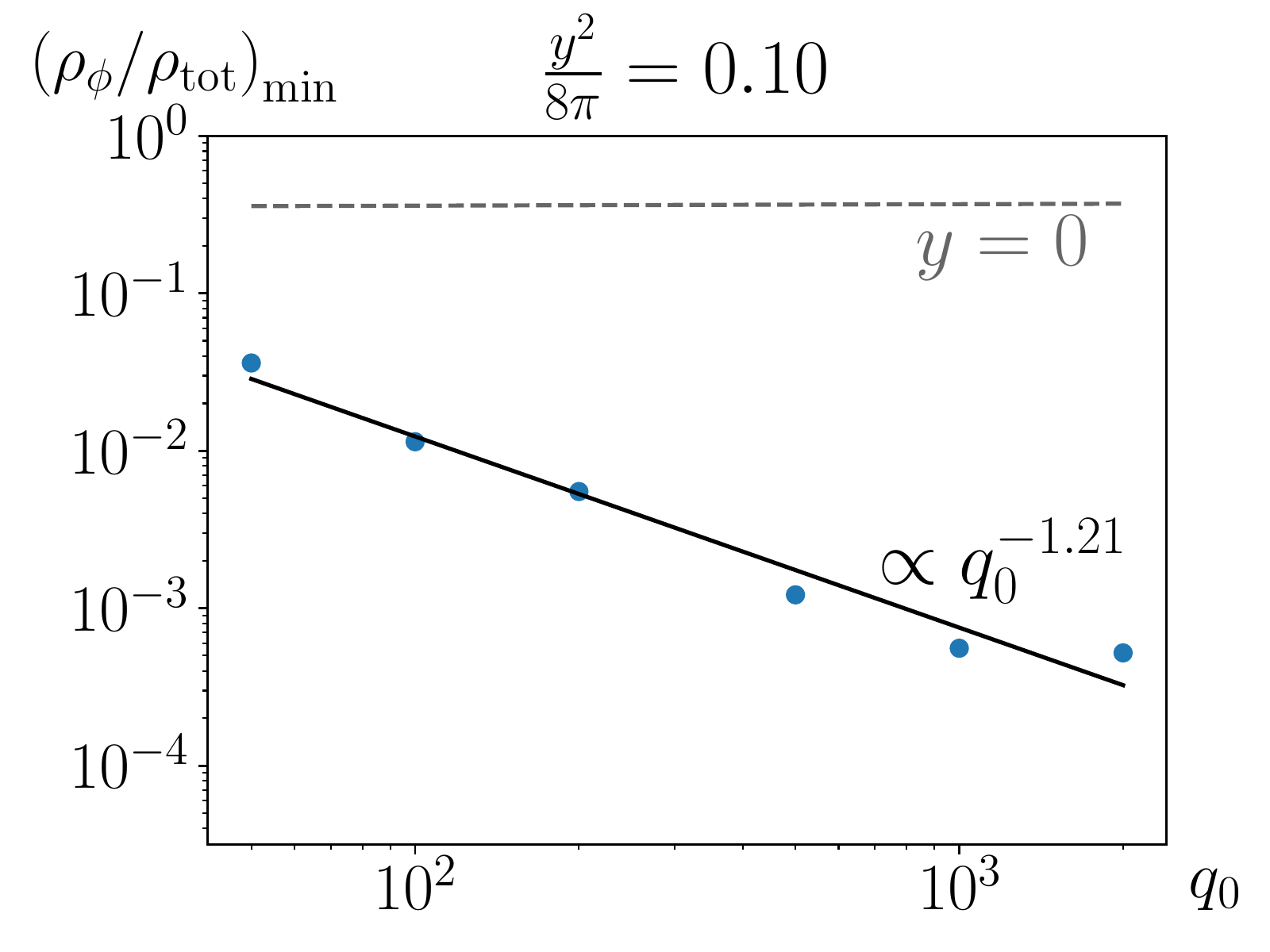}
       \end{subfigure}
       \begin{subfigure}[b]{0.45\textwidth}
            \centering
            \includegraphics[width=\textwidth]{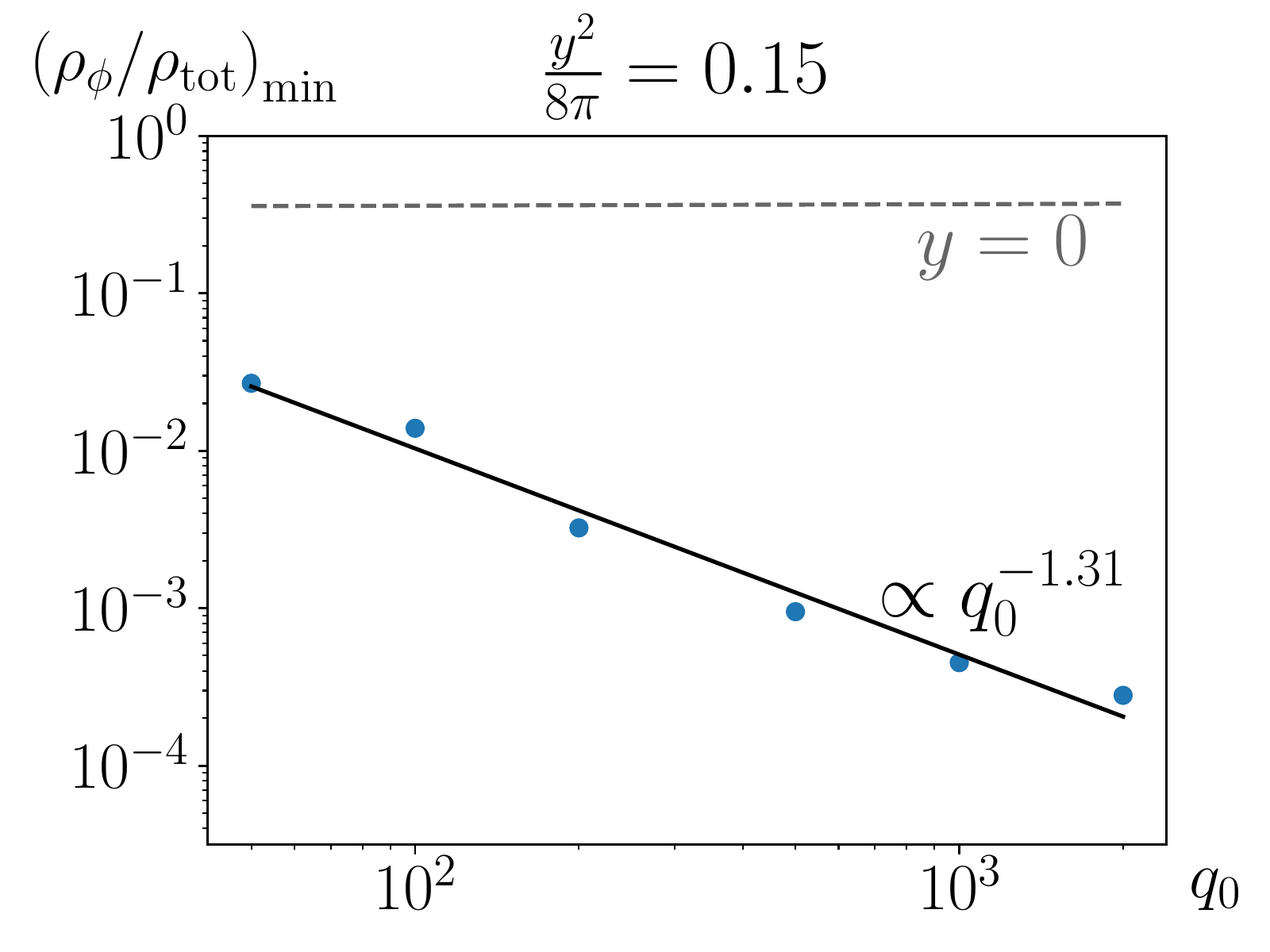}
        \end{subfigure}
    \caption{$(\rho_{\phi}/\rho_{\text{tot}})_{\text{min}}$ as a function of $q_0$ for different choices of $y$. The blue points are the simulation results, and the black line is the best fit with a power law $\propto q_0^{x}$. Each panel also shows in dashed gray the power law best fit for the $y = 0$ case, which is flat at $(\rho_{\phi}/\rho_{\text{tot}})_{\text{min}}\approx 0.5$. We fix $b = 0.9$, $m = 10^{-6} \MP$ and $\Phi_0 = f = \MP$. }
    \label{fig:qscaling}
\end{figure}

This intuition based on the linear analysis is indeed verified by simulation results. We conduct simulations with $q_0=$ 50, 100, 200, 500, 1000, 2000,  and $y^2/(8\pi) =$ 0.01, 0.05, 0.10, 0.15. $m$, $\Phi_0$ and $b$ are fixed at the values specified at the beginning of this section. We use $N = 128$ and $L = 2 m^{-1}$ for all simulations. The $\phi$ and $\chi$ initial power spectra are again cut off at $k_{\phi,\text{max}}/m=0$ and $k_{\chi,\text{max}}/m=2\sqrt{q_0}$. For each parameter choice, the simulation is run for a sufficiently long time until $\rho_{\phi}$ has completely decoupled from $\rho_{\chi}$ and $\rho_{\psi}$, so we can read off the value of $(\rho_{\phi}/\rho_{\text{tot}})_{\text{min}}$.  Fig.~\ref{fig:qscaling} shows how $(\rho_{\phi}/\rho_{\text{tot}})_{\text{min}}$ scales with $q_0$ for different choices of $y$. For every given $y$, we see that $(\rho_{\phi}/\rho_{\text{tot}})_{\text{min}}$ scales with $q_0$ by a simple power law, and is improved by several orders of magnitude compared to the $y = 0$ case which is flat at $(\rho_{\phi}/\rho_{\text{tot}})_{\text{min}}\approx 0.5$. For greater values of $q_0$ beyond our simulation results, we expect the power law improvement to continue. However a definite statement is difficult given the limits of both numerical and analytical understanding of a nonlinear system.


For fixed $q_0$, Fig.~\ref{fig:qscaling} shows that energy transfer efficiency improves (or equivalently, $(\rho_\phi/\rho_{\rm tot})_{\rm min}$ decreases) as $y^2/(8\pi)$ increases. However, we don't expect this improvement to continue to arbitrarily large value of $y$. The energy transfer efficiency should deteriorate when $y\ll 1$ or $y\gg 1$: preheating can only amplify a field value when it is nonzero. When $y$ is so large that $\chi\to\bar{\psi}\psi$ depletes $\chi$ faster than production from preheating, preheating will shut off, and there will be little energy transferred out of the inflaton sector to begin with.\footnote{We note that our classical lattice simulations do not account for the perturbative decay of $\phi$ into pairs of $\chi$. We are allowed to ignore this inherently quantum process here, since it is not efficient during the time interval of our simulation, $\Gamma_{\phi\rightarrow\chi\chi}\sim(M^2/f)^2/m= 2b \lambda m\ll H$, for the parameters chosen here.} On the other hand, when $y \to 0$, the decay $\chi\to\psi$ has little effect on the evolution of the system and we get back to the usual tachyonic resonance scenario, in which $(\rho_\phi/\rho_{\rm tot})_{\rm min} \sim 0.5$ when $q_0 \gg 1$. Given what we expect at the two extreme limits, the energy transfer efficiency must be optimal at some intermediate $y$. The precise optimal value of $y$ could be beyond the range of our simulations.

\begin{figure}
    \begin{subfigure}[b]{0.49\textwidth}
    \centering
    \includegraphics[width=\textwidth]{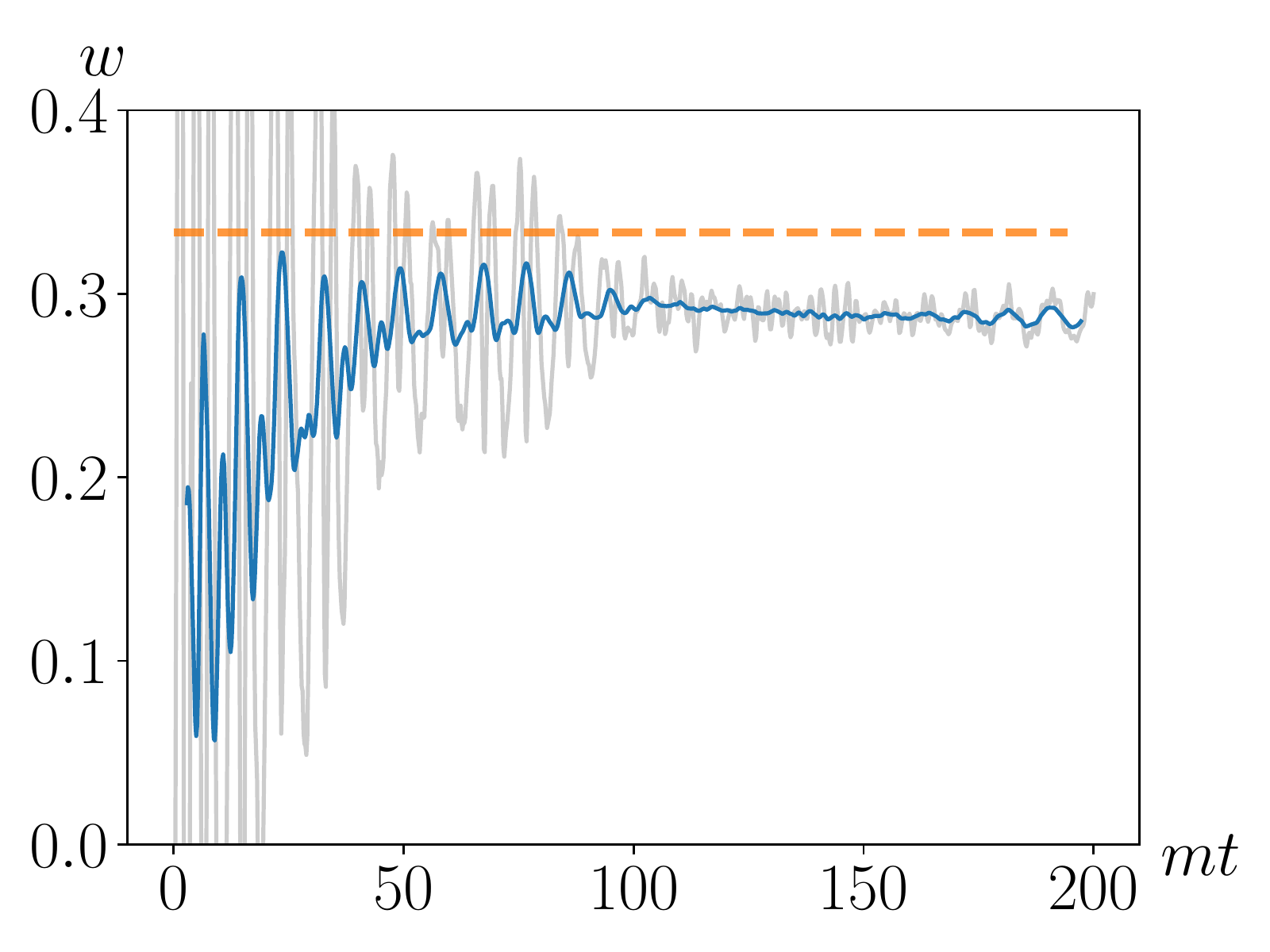}
    \caption{$y^2/8 \pi = 0$}
    \end{subfigure}
    \centering
    \hfill
    \begin{subfigure}[b]{0.49\textwidth}
    \includegraphics[width=\textwidth]{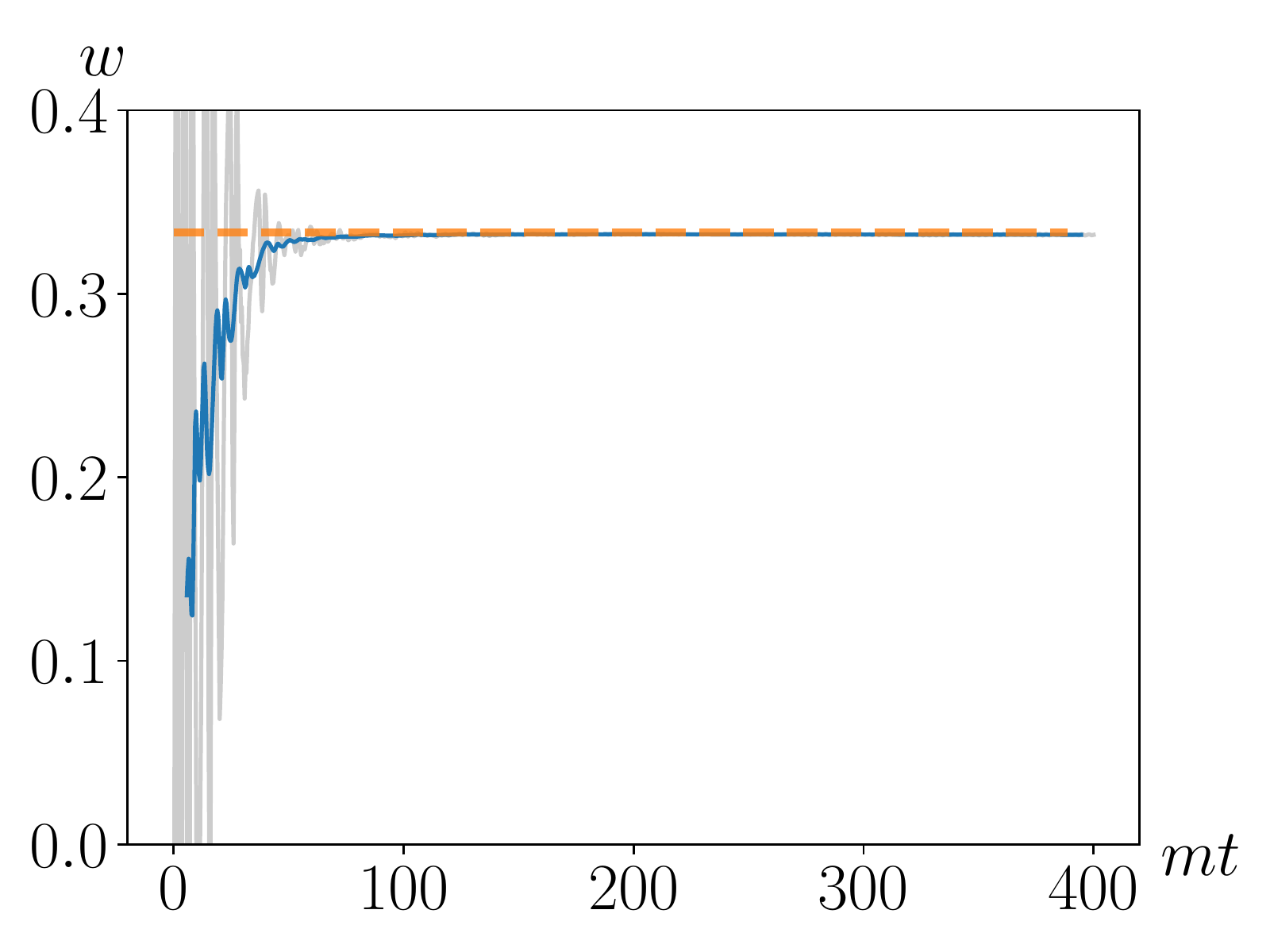}
    \caption{$y^2/8 \pi = 0.1$}
    \end{subfigure}
    \caption{Time evolution of the equation of state $w$ for $b = 0.9$, $m = 10^{-6} \MP$, $\Phi_0 = f = \MP$, $q_0 = 200$ and $y^2/8\pi = 0$ or 0.1. The gray curves are the raw simulation results, and the blue curves are the time average to remove the rapid oscillations. The dashed orange horizontal line is drawn at $w = 1/3$. When $y\neq 0$, the system energy density is dominated by the radiation-like fermion fluid, thus $w$ rapidly approaches $1/3$.}\label{fig:small_lam_w}
\end{figure}

The smaller $(\rho_{\phi}/\rho_{\text{tot}})_{\text{min}}$ is, the more radiation-like the Universe will be at the end of preheating. Fig.~\ref{fig:small_lam_w} shows the time evolution of the equation of state $w$ for systems with $y^2/(8\pi) = 0$ and 0.1. When $y = 0$, there is still a significant energy density left in the inflaton sector. $w$ is in the range between 0 and $1/3$, which corresponds to a mixed matter and radiation state, as reviewed in Sec.~\ref{subsec:tach}. When $y\neq 0$,  the efficiency to transfer energy from inflaton to radiation is improved dramatically, and $w$ converges rapidly to $1/3$. While there is still a tiny fraction of energy density left in inflaton, it takes much longer for the inflaton to dominate the energy density again due to the slower redshift. 

\begin{figure}[t]
    \centering
    \begin{subfigure}[b]{\textwidth}
         \centering
         \includegraphics[width=0.49\textwidth]{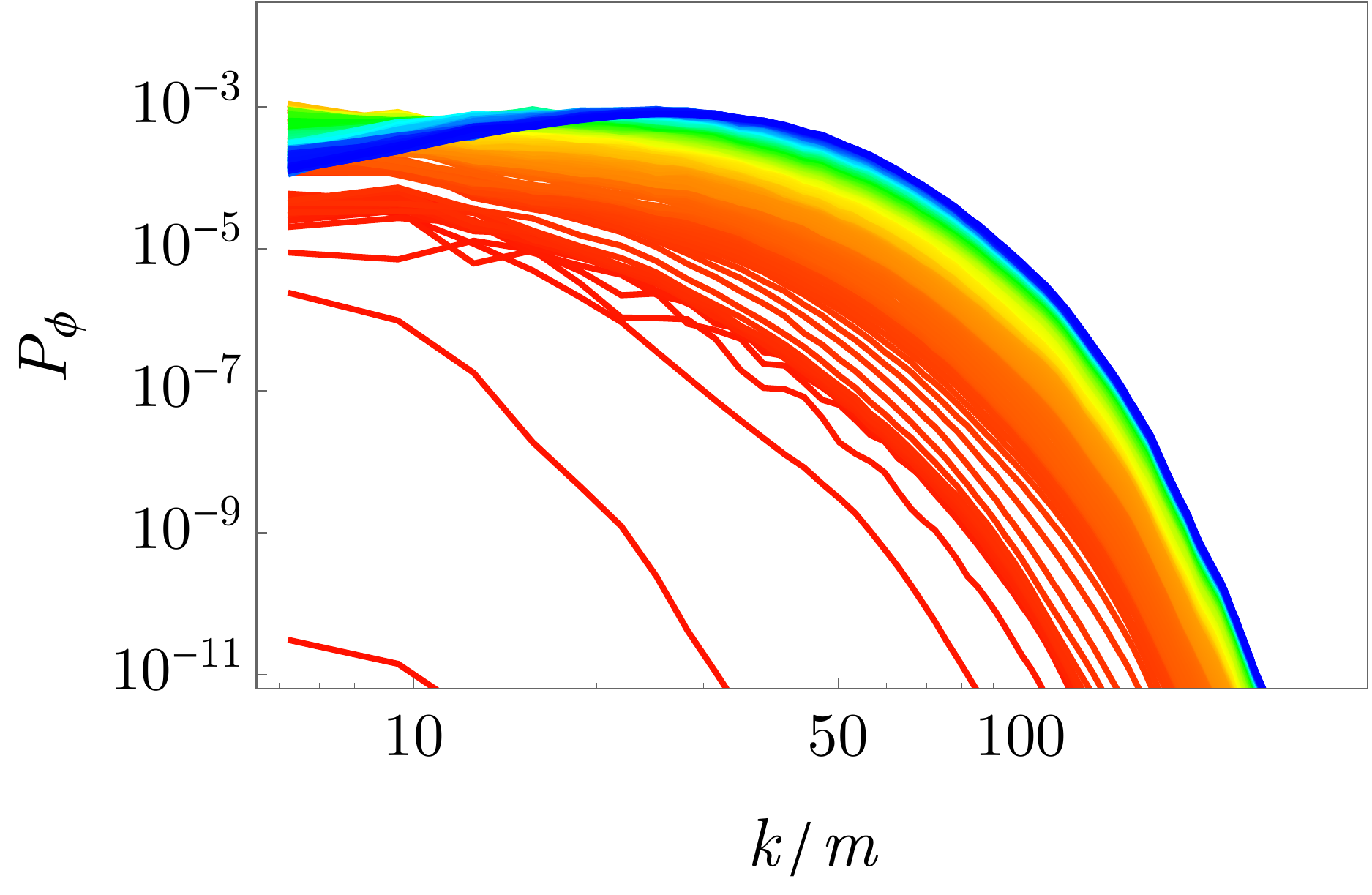}
          \includegraphics[width=0.49\textwidth]{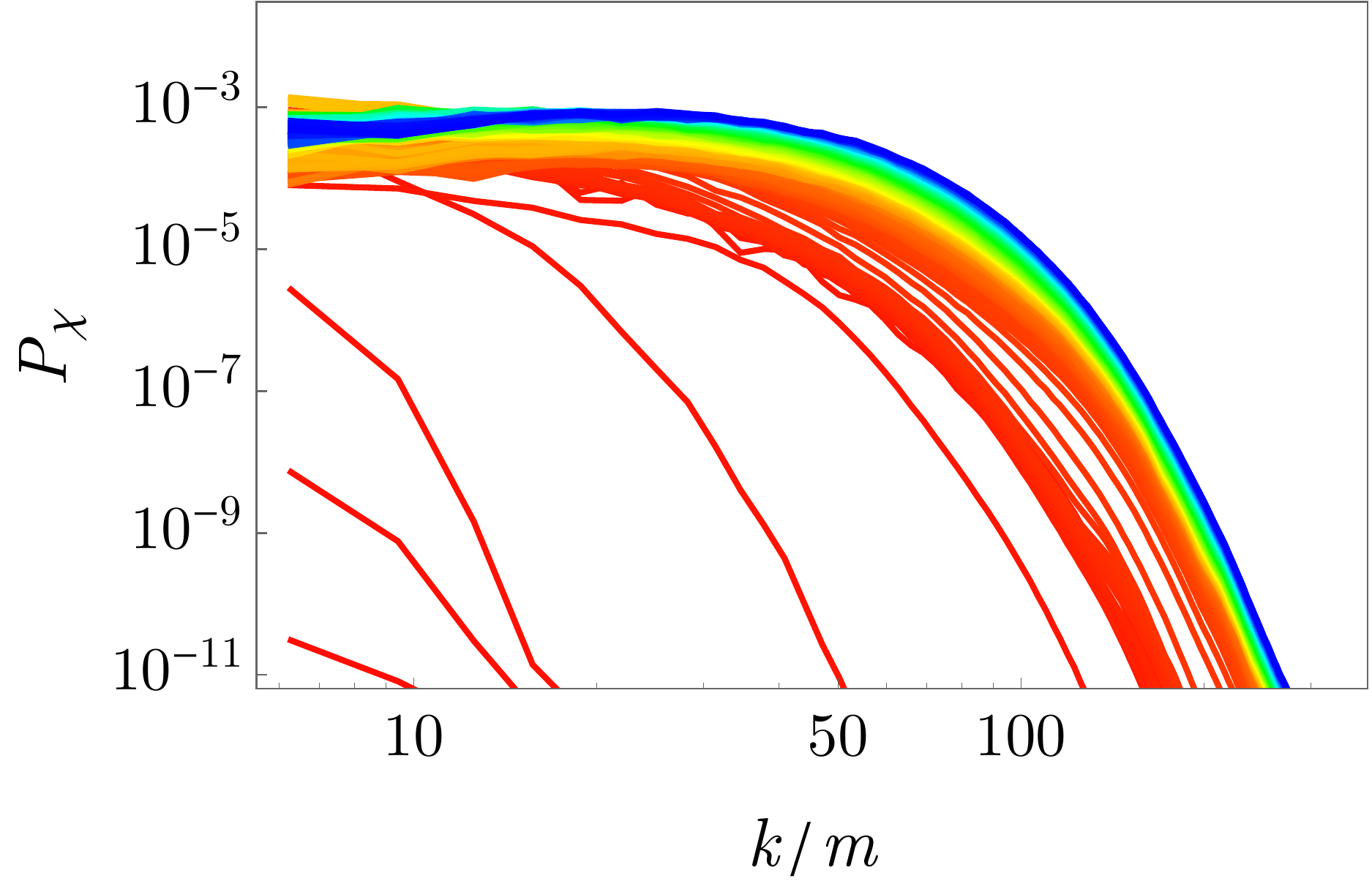}
      \caption{$y^2/8\pi = 0$}
      \end{subfigure}
      \par\bigskip
      \begin{subfigure}[b]{\textwidth}
           \centering
           \includegraphics[width=0.49\textwidth]{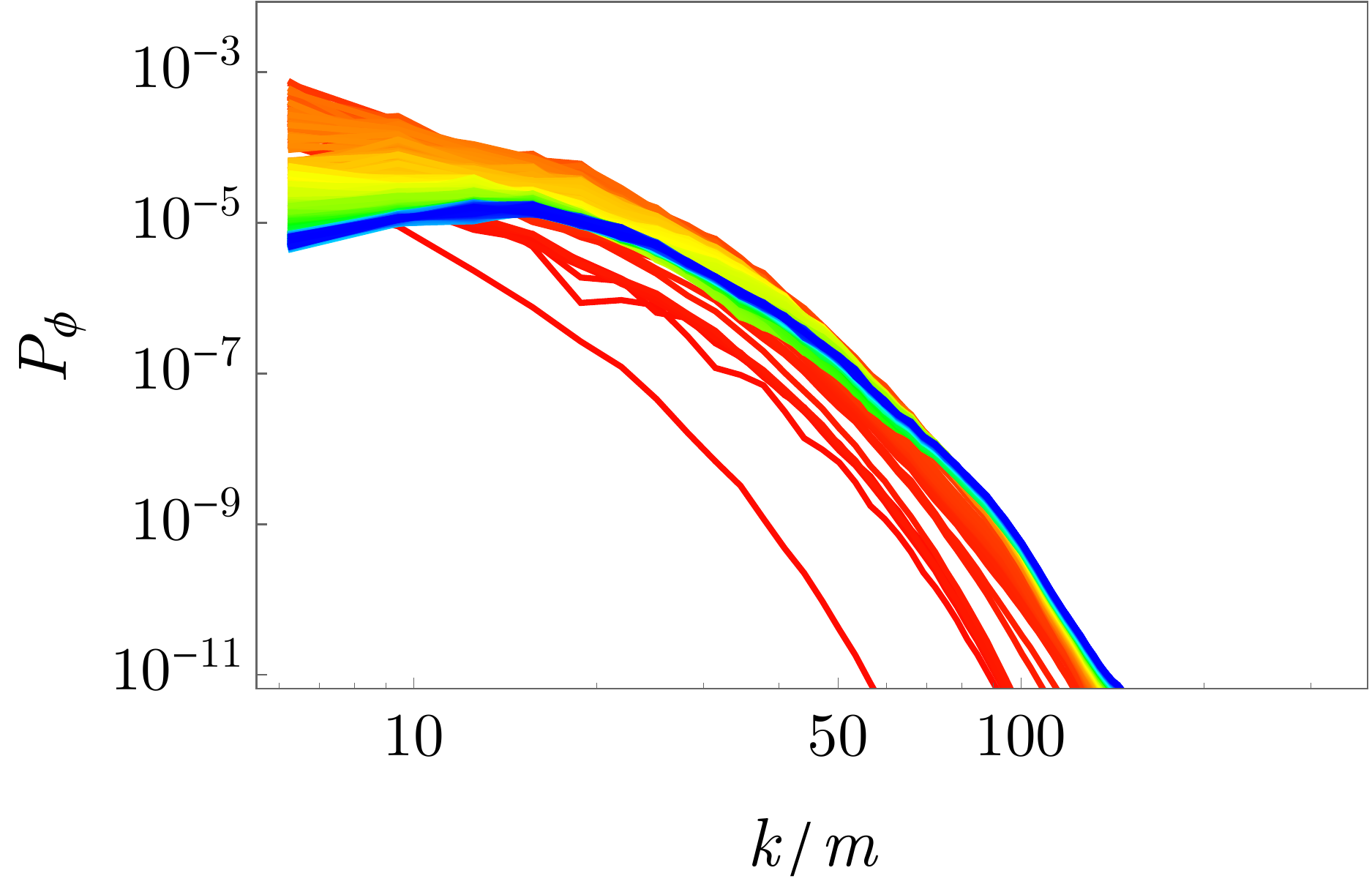}
            \includegraphics[width=0.49\textwidth]{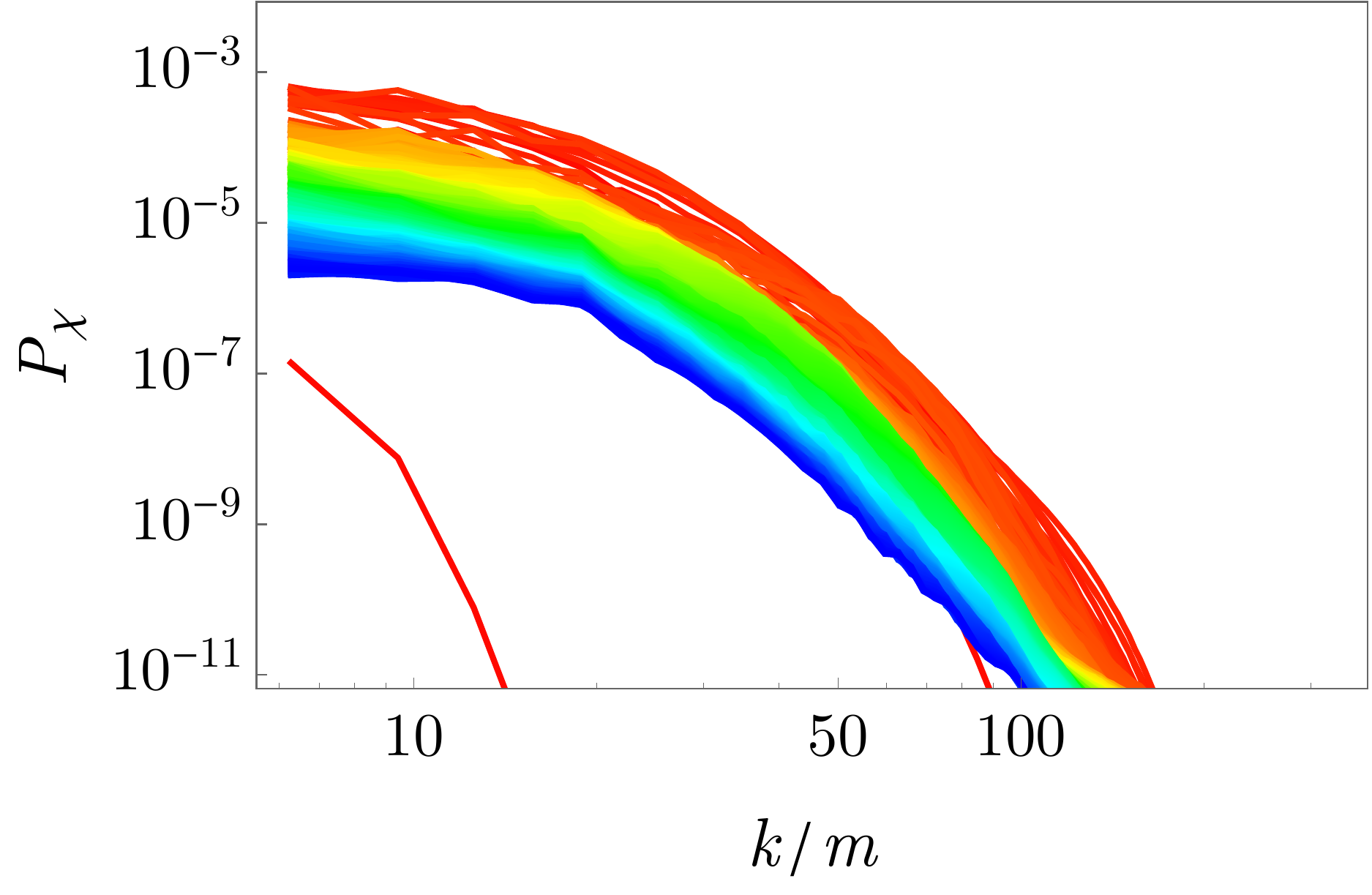}
            \caption{$y^2/8\pi = 0.1$}
        \end{subfigure}
    \caption{The time evolution of field power spectra for $b = 0.9$, $m = 10^{-6} \MP$, $\Phi_0 = f = \MP$, $q_0 = 200$, and $y^2/8\pi = 0$ or 0.1. For each field $F$, $P_F$ is the comoving power $P_F\equiv (\dd/\dd\ln k)\overline{a^3F(x)^2}$ in units of $\MP$. Time evolves from red to blue. For $y^2/(8\pi)=0.1$, there is little propagation of power to the UV compared to the $y = 0$ case.}
    \label{fig:small_lam_specs}
\end{figure}

Another distinctive feature of the system with perturbative decays is that in the nonlinear stage, the system with perturbative decays has much slower power propagation to the UV end of the spectra. \Fig{fig:small_lam_specs} shows the time evolution of the power spectra of $\phi$ and $\chi$ for $q_0 = 200$ and $y^2/(8\pi) = 0$ or $0.1$. For both $y^2/(8\pi) = 0$ and 0.1, there is an initial exponential growth in the power of $\chi$ due to tachyonic resonance. For $y=0$, the system quickly enters the nonlinear stage with $\rho_{\chi}\approx \rho_{\phi}$, and power spectra keep growing due to rescattering. Higher $k$ modes are excited and power gradually propagates to the UV end. However, for $y^2/(8\pi)=0.1$, $\phi$ is much more depleted than in the $y^2/(8\pi)=0$ case, hence there is little backreaction/rescattering between $\phi$ and $\chi$ and less power is transferred to the UV. The power spectrum of $\phi$ is mostly undisturbed at the later stage of the evolution, while the power spectrum of $\chi$ decreases in magnitude due to the perturbative decays.

\subsection{Constraints on the parameters}
\label{subsec:backreaction}

Results presented in the last section show that having perturbative decays of $\chi$ can greatly reduce $(\rho_{\phi}/\rho_{\text{tot}})_{\text{min}}$ and improve the efficiency to transfer the inflaton energy density to radiation. However, there are two important implicit assumptions in our simulation. Firstly, we ignore the backreaction of the fermionic fluid to the scalar system and treat the fluid as an infinite energy sink. Secondly, we ignore Pauli repulsion effects which can potentially prohibit the decays $\chi\to\psi$ from happening. In this section, we present analytical arguments on the parametric relations required for these assumptions to be valid. 

We start with estimating the fermionic backreaction. There are two possible ways to derive the conditions when the backreaction becomes non-negligible or vice versa. The first way is to check the effective fermion mass generated by a nonzero value of $\chi$. In our simulations, we take the fermions to be massless by modeling them as a radiation-like fluid. Yet the Yukawa coupling, $y \chi \bar{\psi} \psi$, generates an effective fermion mass:
\beq
m_{\psi}\sim y\expval{|\chi|}\sim y \sqrt{\frac{M^2|\Phi|}{\lambda f}} \sim y\frac{M}{\sqrt{\lambda}},
\eeq
where $\expval{|\chi|}$ is set to be its value at the minimum of the potential when $\phi$ oscillates to the tachyonic side. We ignore the time evolution of $\Phi$ in the estimate above. For the decays $\chi \to \bar{\psi}\psi$ to happen, we need $E_{\chi}\gtrsim m_{\psi}$. Both analytical and numerical analyses show that the characteristic value of $\chi$'s energy is $E_{\chi}\sim M$. Therefore, the kinematic constraint, $E_{\chi}\gtrsim m_{\psi}$, translates into
\beq
\frac{y}{\sqrt{\lambda}}\lesssim 1.\label{eq:deri1}
\eeq
This is a rough upper bound since we ignore the time evolution of relevant quantities. 

Once there is a large number of fermions around, they will induce a tadpole term to the potential of $\chi$, $y \chi \langle \bar{\psi} \psi \rangle$.   
Thus the other way to check the importance of backreaction is to compare the tadpole term with the other terms, i.e., the term that drives the tachyonic resonance production, $(M^2/f) \phi \chi^2$, in the scalar potential. When the backreaction is sub-dominant, we should have 
\beq
y\chi\expval{\bar{\psi}\psi}\lesssim \frac{M^2}{f}\phi\chi^2.
\eeq
The fermion condensate can be approximated as \beq
\expval{\bar{\psi}\psi}\sim \frac{\rho_{\psi}}{E_{\psi}}\sim \frac{m^2f^2}{M},
\eeq
where we estimate $\rho_{\psi}$ to be the initial energy density of $\phi$ since we expect most of the energy is transferred from $\phi$ to $\psi$ eventually. We also approximate $E_{\psi}\sim E_{\chi}\sim M$, $\phi \sim f$, and $\chi^2 \sim M^2/\lambda$. The condition becomes
\beq
y\frac{m^2f^2}{M}\lesssim \frac{M^3}{\sqrt{\lambda}} \quad \Rightarrow \quad \frac{y}{\sqrt{\lambda}}\lesssim 1, \label{eq:deri2}
\eeq
where in the last step, we use the fact that for efficient tachyonic resonance production, $b = M^4/(2 \lambda m^2 f^2)$ has to be close to one. Note that this is the same as Eq.~\eqref{eq:deri1}. Both arguments from somewhat different points of view lead to the same parametric relation for the fermion's backreaction to be negligible.

Now we consider how to avoid Pauli blocking. Pauli blocking will prevent $\chi\to\psi$ decays if the phase space is not big enough to accomodate the fermionic decay products. To check that, we need to estimate the occupation number of $\psi$. The number density of $\psi$, when the fermion fluid becomes the dominant component of the energy density, could be estimated as 
\beq
n_{\psi}\sim \frac{\rho_{\psi}}{E_{\psi}}\sim \frac{m^2f^2}{M}. 
\eeq
The occupation number of $\psi$ is then 
\beq
\frac{n_{\psi}}{k_{\psi}^3} \sim \frac{m^2f^2}{M^4} \sim \frac{1}{\lambda},
\eeq
where the range of momenta of $\psi$, $k_\psi$, is set by the typical energy of $\chi$ produced from tachyonic resonance, $k_{\chi}\sim M$. In the last step, we again use the fact that $b$ has to be close to one. 

At first glance, in order for the occupation number of $\psi$ to be smaller than 2 so that the decays are free from Pauli blocking, we simply need $\lambda$ to be of order one. Yet this choice is problematic. The phase space volume of $\psi$ is similar to that of $\chi$ and the occupation number of $\chi$ could be estimated in the same way to be $1/\lambda$. An order one $\lambda$ then implies that $\chi$ is on the border line of being treated as a classical field. More importantly, the perturbative decays of inflaton through the trilinear coupling $M^2\phi \chi^2/f$ happens on the time scale of
\beq
\Gamma^{-1}_\phi \sim \frac{8 \pi f^2m}{M^4} \sim \frac{8\pi}{\lambda m}.
\eeq
Since the preheating happens on the time scale of ${\cal O} (1-10) m^{-1}$, an order one $\lambda$ indicates that the time scales of perturbative reheating and preheating are about the same and there is no need to consider preheating in this case. 
Thus in the parameter space where preheating matters and occurs before reheating, $\lambda \ll 1$. In order to enhance the energy transfer efficiency without Pauli blocking in our model, we need to introduce $N_f$ species of fermions, $\psi_{i = 1\cdots N_f}$. For simplicity, we assume that the Yukawa couplings of all the fermions have the same value, $y_i = y'$. The occupation number of each fermion species is then $1/(N_f \lambda)$, which tells us to circumvent Pauli blocking,
\beq
N_f \lambda \gtrsim 1.
\eeq
While too many fermion species imply a too low cutoff of our toy model as a valid effective field theory, it is reasonable to consider a (not too) small $\lambda$, e.g., $\lambda \sim {\cal O}(0.01)$ and $N_f \sim {\cal O} (100)$. 

From the simulation point of view, it makes no difference implementing one fermion fluid or many fermion fluids, if we do not explicitly include Pauli blocking terms. For simplicity, we proceed with simulations with a single fermion fluid, with the implicit understanding that this is an approximation of an $N_f$-fermion system with $N_f\gtrsim 1/\lambda$. $\rho_{\psi}$ is then the total energy density of the $N_f$ fermions, and the decay width $\Gamma_{\chi} = y^2m_{\chi}/(8\pi^2)$ is the sum of the decay widths to individual fermions, $\Gamma_{\chi\to \psi_i} = y'^2m_{\chi}/(8\pi^2)$. In other words, the Yukawa couplings are related by $y^2=N_f y'^2$. When we include multiple fermions in the system, the condition for negligible fermion backreaction in Eq.~\eqref{eq:deri1} and Eq.~\eqref{eq:deri2} should be understood as constraints on $y'$, the coupling of $\chi$ to an individual fermion,
\beq
y'^2 = \frac{y^2}{N_f} \lesssim \lambda\Rightarrow y^2 \lesssim N_f\lambda.
\eeq

Combining the analytic estimates above with results in the previous section, we find that in order to make the spillway preheating a much more efficient preheating mechanism, there are multiple requirements on the parameters involved:
\begin{enumerate}
    \item Have efficient tachyonic resonance production: $q_0 \gg 1$ and $b \sim 1$ or equivalently $M \gg m$ and $M^2 \sim \sqrt{\lambda} m f$.
    \item Tachyonic resonance is the dominant mechanism of energy transfer $\phi \to \chi$, or equivalently perturbative decays of $\phi$ is inefficient during the preheating stage: $\lambda \ll 1$.
    \item Perturbative decays of $\chi$ happen around the time when the energy transferred to $\chi$ through tachyonic particle production becomes comparable to $\rho_\phi$: $\Gamma_\chi^{-1} \sim {\cal O} (1-10) m^{-1}$. Equivalently, $y^2/(8\pi)= N_f y'^2/(8\pi) \sim m/M$ up to some order one numerical factor.
    \item Satisfy the CMB constraint, i.e., the normalization of the scalar perturbation, on the inflaton mass scale: $m \sim 10^{-6}\MP$ for quadratic chaotic inflation.
    \item Free from Pauli blocking of fermions: there are $N_f$ species of fermions with similar Yukawa couplings and $N_f \gtrsim 1/\lambda$. 
    \item Free from backreaction of fermions: $y'^2=y^2/N_f \lesssim \lambda$.
    \end{enumerate}

The system that satisfies all the requirements would have 
\beq
m \sim 10^{-6}\MP, \; \; m \ll M \ll f \sim \MP, \;\;  y^2 \sim 8\pi \frac{m}{M} \lesssim N_f\lambda, \;\;\sqrt{mf} \gg M \gtrsim \left(m^3f^2/N_f\right)^{1/5}, \;\; N_f \gtrsim \frac{1}{\lambda}.
\label{eq:ideal}
\eeq
The simulations shown in the previous section satisfy all conditions, as long as $N_f\gtrsim 1/\lambda \sim 10^{6}-10^{9}$, where system with a smaller $q_0$ require a larger $N_f$. Given that the required value of $N_f$ is large, we need to make sure that the cutoff of our effective field theory (EFT) is not too low: the scale at which gravity becomes strongly coupled and the EFT description breaks down is $M_{\text{pl,eff}} \sim M_{\text{pl}}/\sqrt{N_f}\sim (10^{-4.5}-10^{-3})M_{\text{pl}}$ \cite{Dvali:2007wp}. The maximum comoving momentum excited in the simulations is typically on the same order as that shown in \Fig{fig:small_lam_specs}, with $k\sim 100 m = 10^{-4} M_{\text{pl}}$. Taking into account the expansion of the universe, the physical momentum excited in the system is $k_{\text{phys}}\sim k/a\sim 10^{-5} M_{\text{pl}}$. This is somewhat close to the gravitational cutoff, but smaller, so the EFT description is still safe.

System with a smaller $N_f$ (thus larger $\lambda$) with the same $m$ and $f$ would require a larger $M$, which is computationally more expensive to simulate. The maximum $k$-mode excited by the tachyonic resonance scales as $k_{\text{max}}/m = \sqrt{q_0} =M/m$, and the number of gridpoints required to cover such a $k$-range increases as $N \sim k_{\max}/m$. CPU-time needed grows at least as $N^3$, and potentially more because higher $k$ modes require smaller time steps to resolve.

However, system with a smaller $N_f$ is numerically feasible in an alternative part of the parameter space, where the three mass scales $m$, $M$, $f$ are close to each other. In the next section, we will consider $f \sim {\cal O}(10) M$ and $M \sim {\cal{O}}(10) m$. This allows $y \sim\lambda\sim \mathcal{O}(1)$ without making the simulation computationally infeasible. System with this choice of parameters violates the second requirement (slow perturbative decay of $\phi$) and fourth requirement (CMB constraint), but the essential features of the spillway preheating mechanism is intact. Moreover, the minimum required $N_f$ is $N_f\gtrsim 1/\lambda \sim \mathcal{O}(1)$, much smaller compared to what is needed the previous section. This will be an independent check of the results obtained in the previous section in a qualitatively different region of the parameter space.

\subsection{Alternative simulations }
\label{subsec:alternative}
We consider a suite of alternative simulations based on the following parameters
\beq
f =\MP,\quad m =1.3\times 10^{-2} \MP,\quad q_0= \frac{M^2}{m^2} = 200, \quad {\rm and} \quad b = 0.9,
\eeq
which corresponds to $\lambda = 4.05$. 
As in \Sec{subsec:enhanced}, we simulate the system on a box of length $L =2 m^{-1}$ with $128^3$ points and $y^2/(8\pi) = 0$ and 0.1. We also put a UV cutoff on the initial power spectra of $\phi$ and $\chi$ at $k_{\phi,\text{max}}/m = 0$ and for $\chi$ we cut off at $k_{\chi,\text{max}}/m = 2\sqrt{q_0}$.\footnote{The default initial field fluctuations set by LatticeEasy makes $\rho_{\phi}(t=0)\approx \rho_{\chi}(t=0)$. We manually decrease the magnitude of the initial fluctuations of $\chi$ by a factor of $10^3$ to make $\rho_{\chi}(t=0)\ll \rho_{\phi}(t=0)$, so that it is easier to observe the interplay between tachyonic resonance and the $\chi\rightarrow \psi\psi$ decays.}

\begin{figure}[h]
    \begin{subfigure}[b]{0.45\textwidth}
    \centering
    \includegraphics[width=\textwidth]{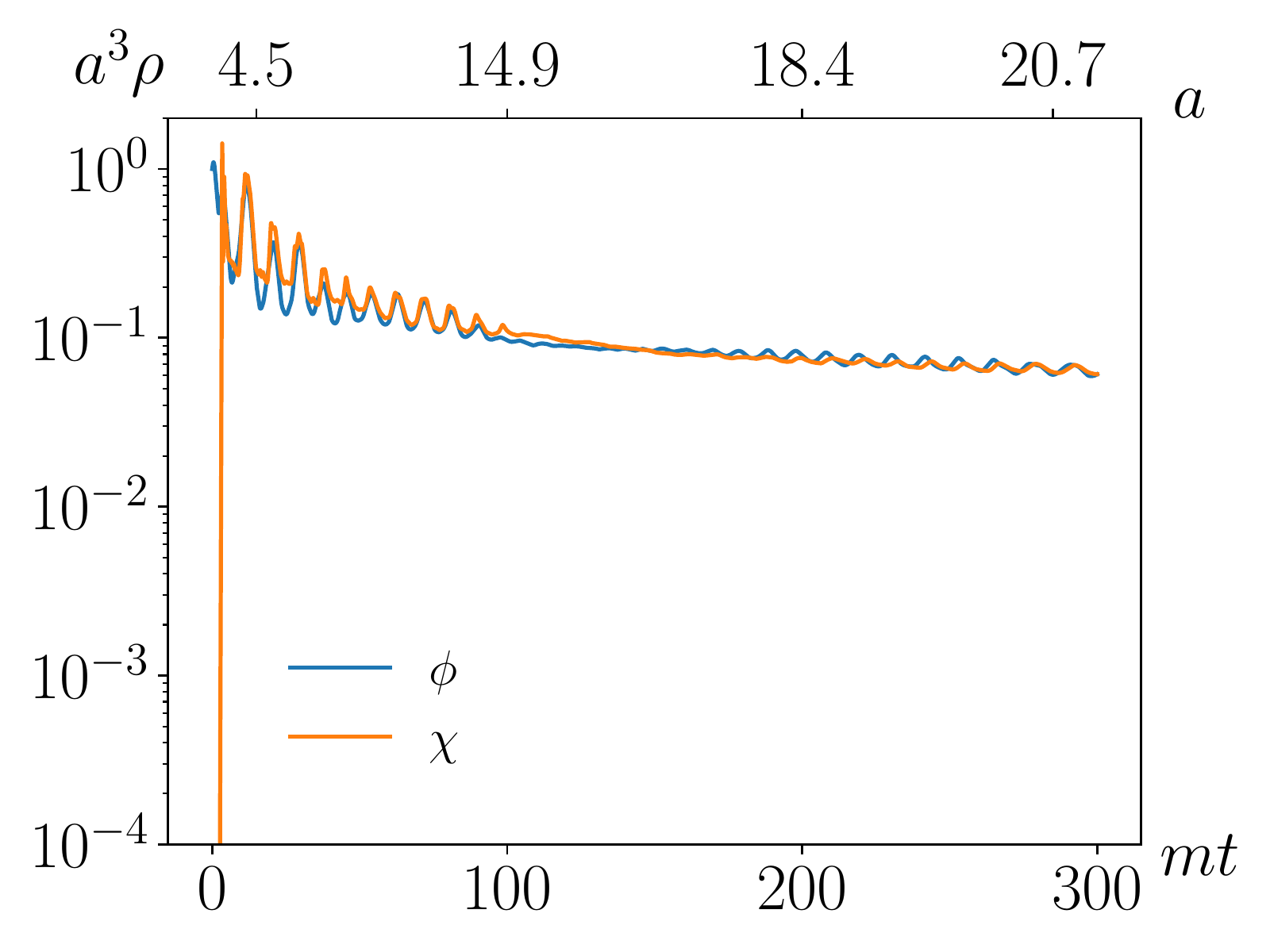}
    \caption{$y^2/8 \pi = 0$}
    \end{subfigure}
    \centering
    \begin{subfigure}[b]{0.45\textwidth}
    \centering
    \includegraphics[width=\textwidth]{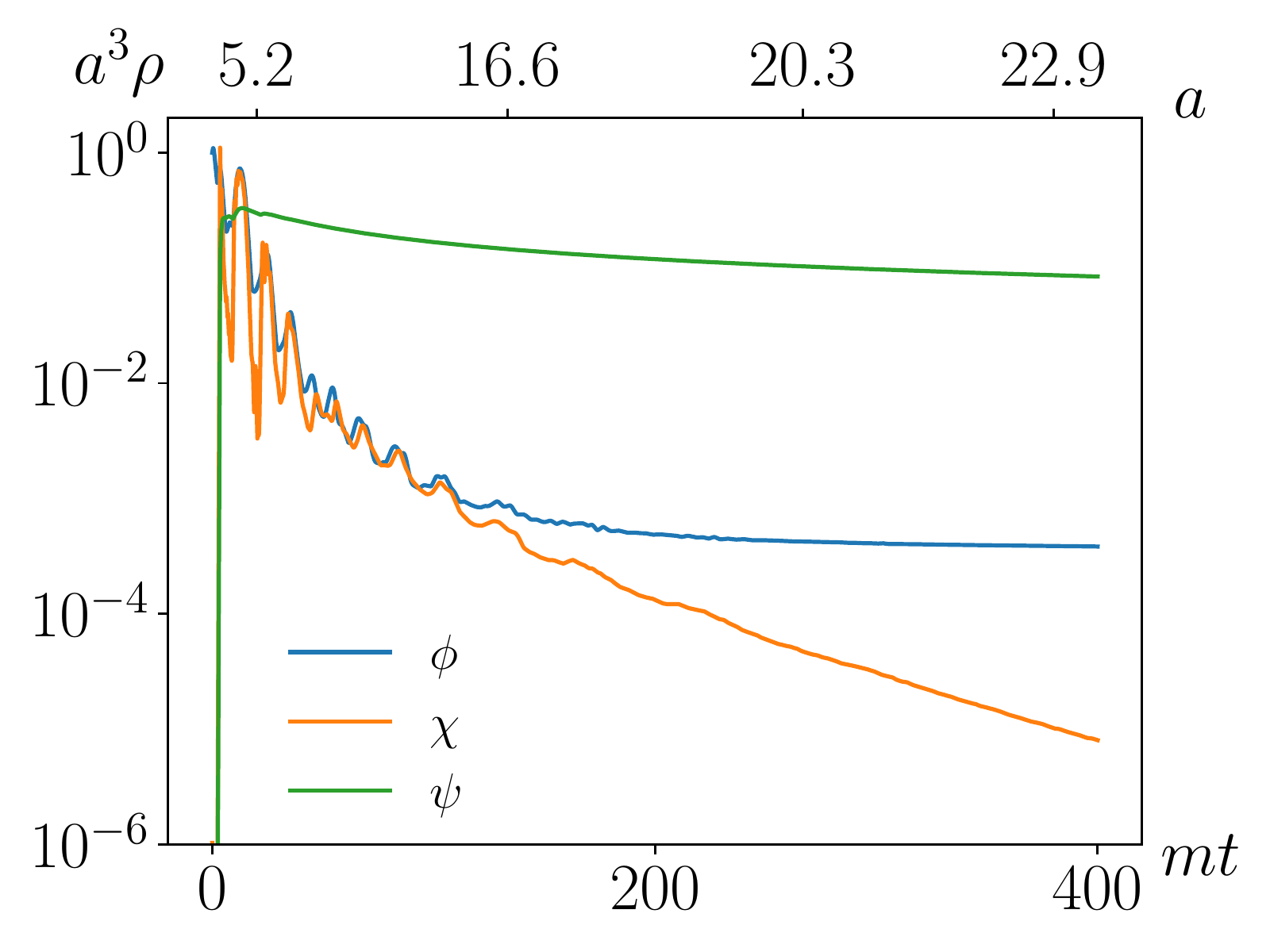}
    \caption{$y^2/8 \pi = 0.1$}
    \end{subfigure}
    \hfill
    \caption{Time evolution of $\phi$, $\chi$, and fermion fluid energy density for $b = 0.9$, $m = 1.3\times 10^{-2} \MP$, $\Phi_0 = f = \MP$, $q_0 = 200$ and $y^2/8\pi = 0$ or 0.1.}\label{fig:large_lam_rhoEvol}
\end{figure}

The time evolution of the system with either $y^2/(8\pi) = 0$ or $0.1$ is shown in \Fig{fig:large_lam_rhoEvol}. Apart from short-term oscillations of the energy densities, the system evolution for both $y = 0$ and $y^2/(8\pi) = 0.1$ is qualitatively similar to what we have observed in \Sec{subsec:enhanced}. When $y = 0$, $\rho_{\chi}$ quickly builds up due to tachyonic resonance. But once $\rho_{\chi}\approx\rho_{\phi}$, their ratio stays constant for a long time. For $y^2/(8\pi) = 0.1$, there is a significant enhancement of energy transfer out of the inflaton due to the $\phi\to\chi\to\psi$ cascade decays.

\begin{figure}[h]
    \centering
    \begin{subfigure}[b]{0.45\textwidth}
    \centering
    \includegraphics[width=\textwidth]{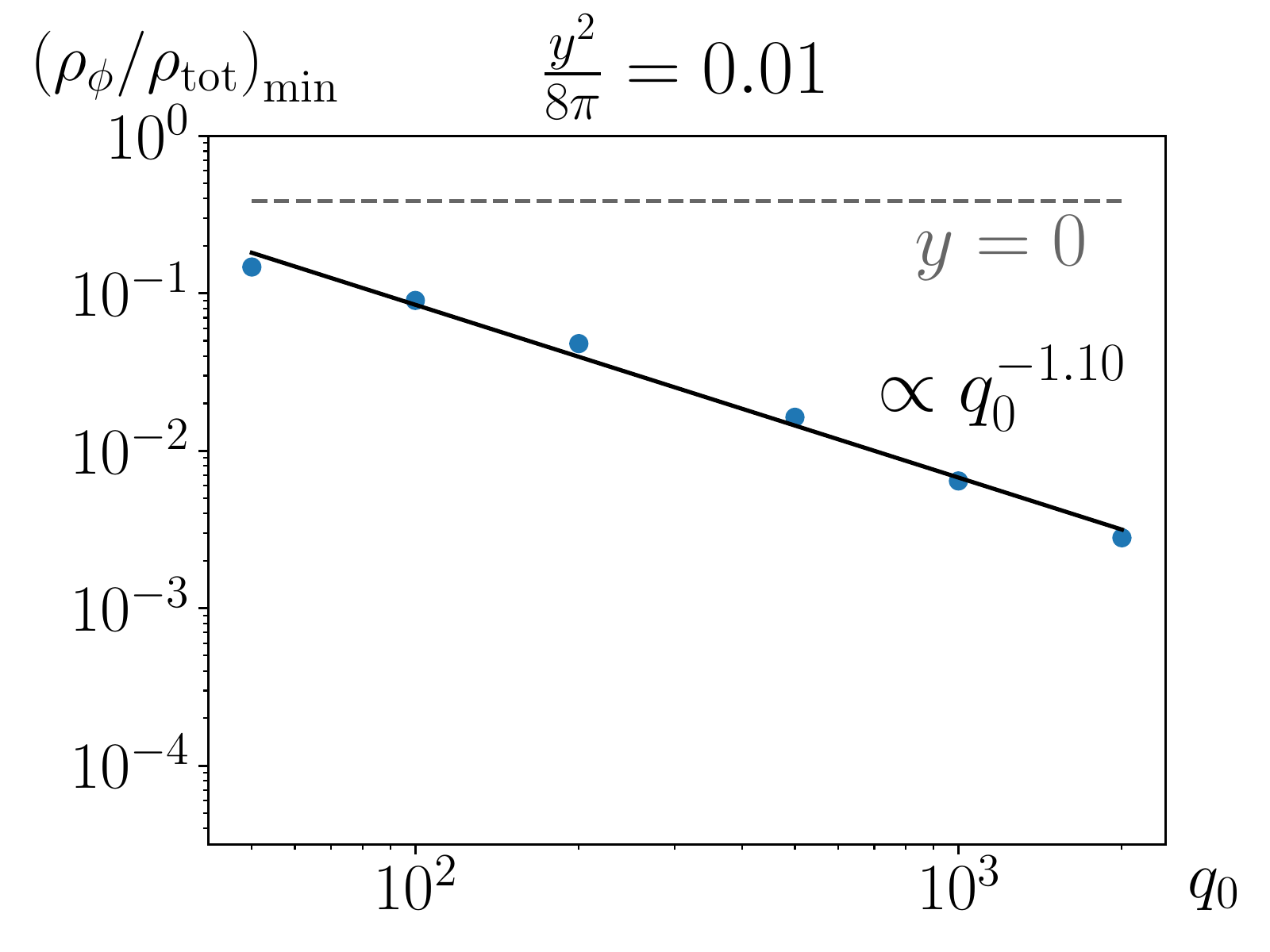}
    \end{subfigure} 
    \centering
    \begin{subfigure}[b]{0.45\textwidth}
    \centering
    \includegraphics[width=\textwidth]{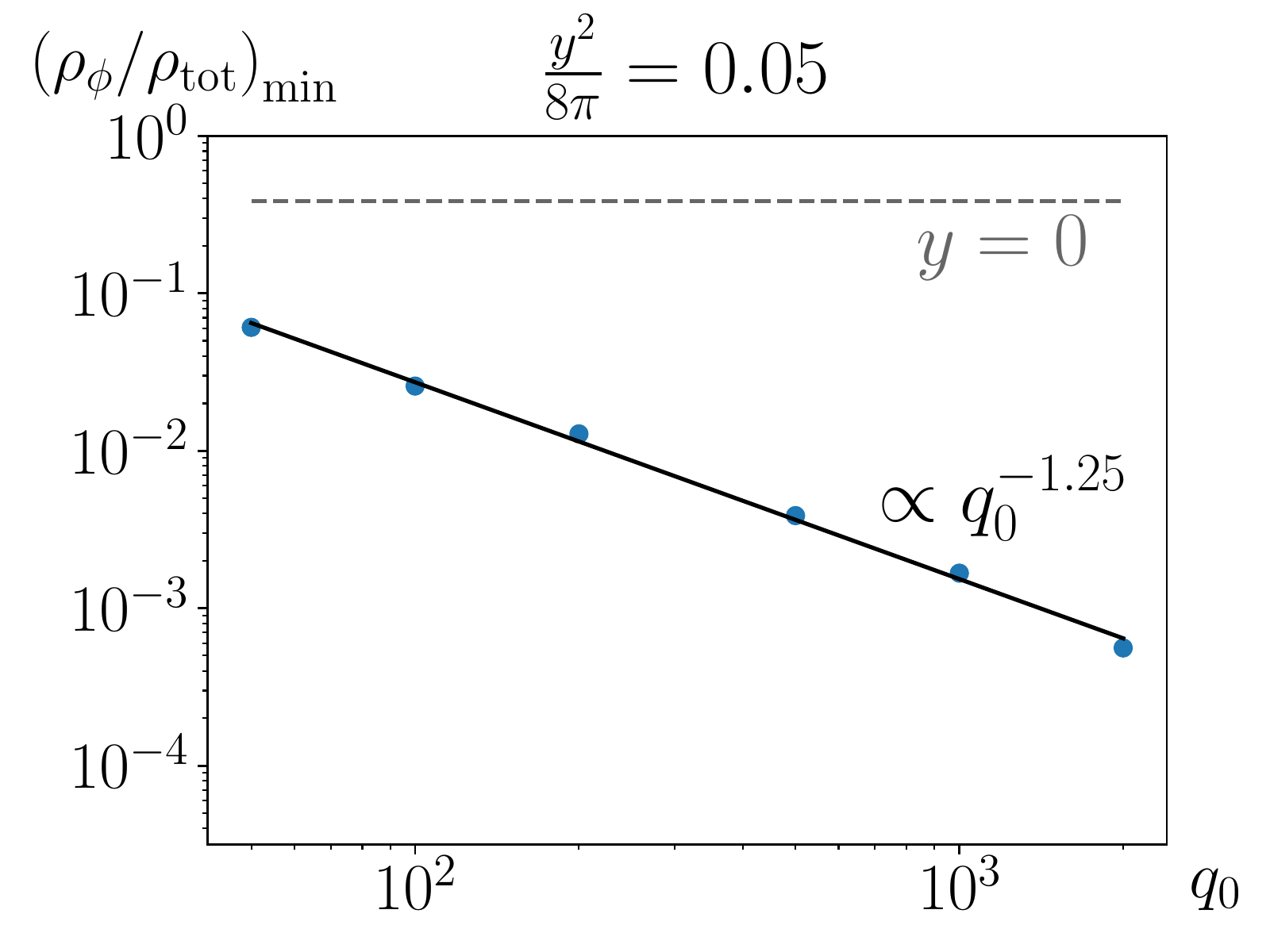}
    \end{subfigure}
    \centering
    \begin{subfigure}[b]{0.45\textwidth}
    \centering
    \includegraphics[width=\textwidth]{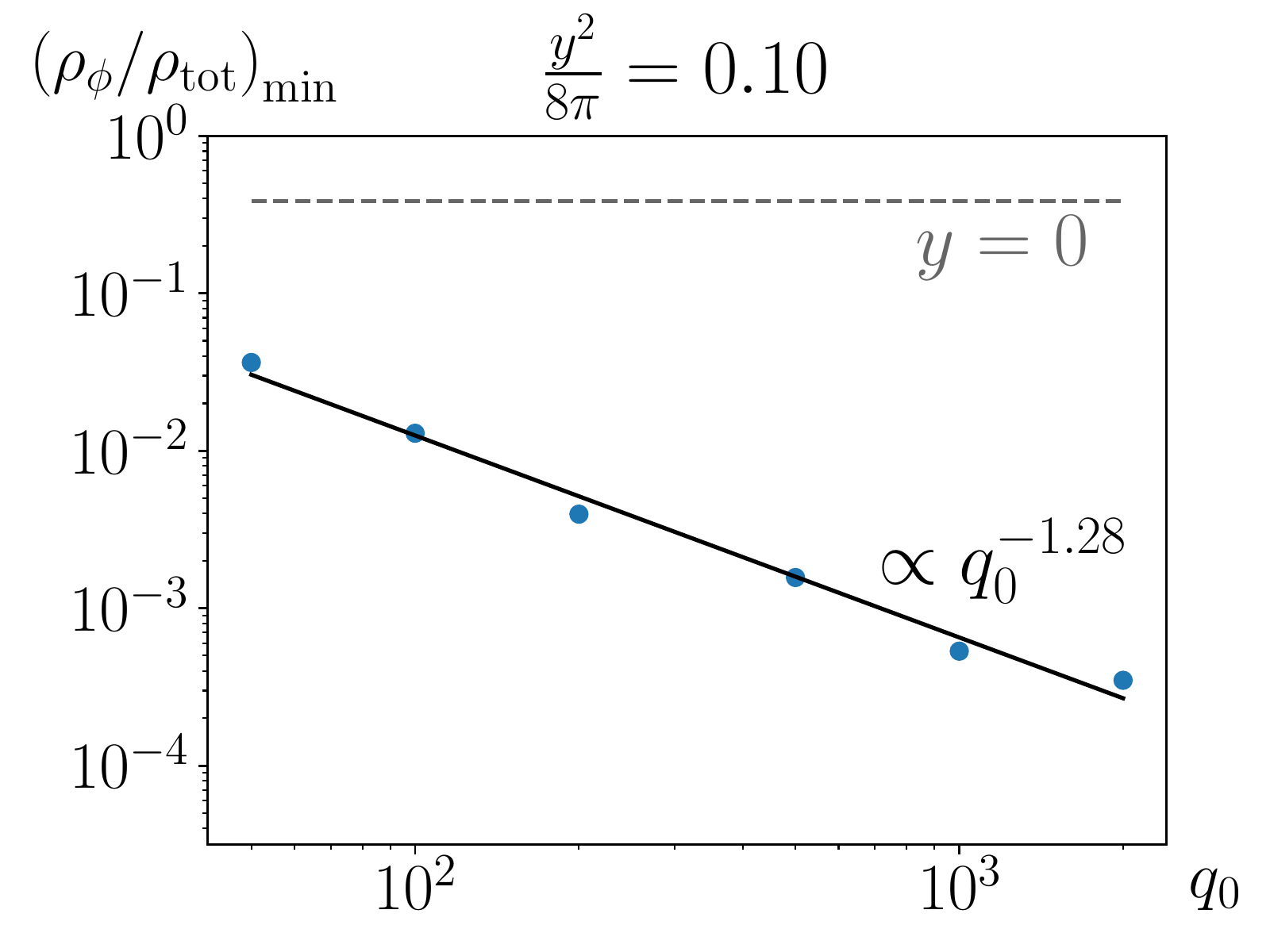}
    \end{subfigure}
    \centering
    \begin{subfigure}[b]{0.45\textwidth}
    \centering
    \includegraphics[width=\textwidth]{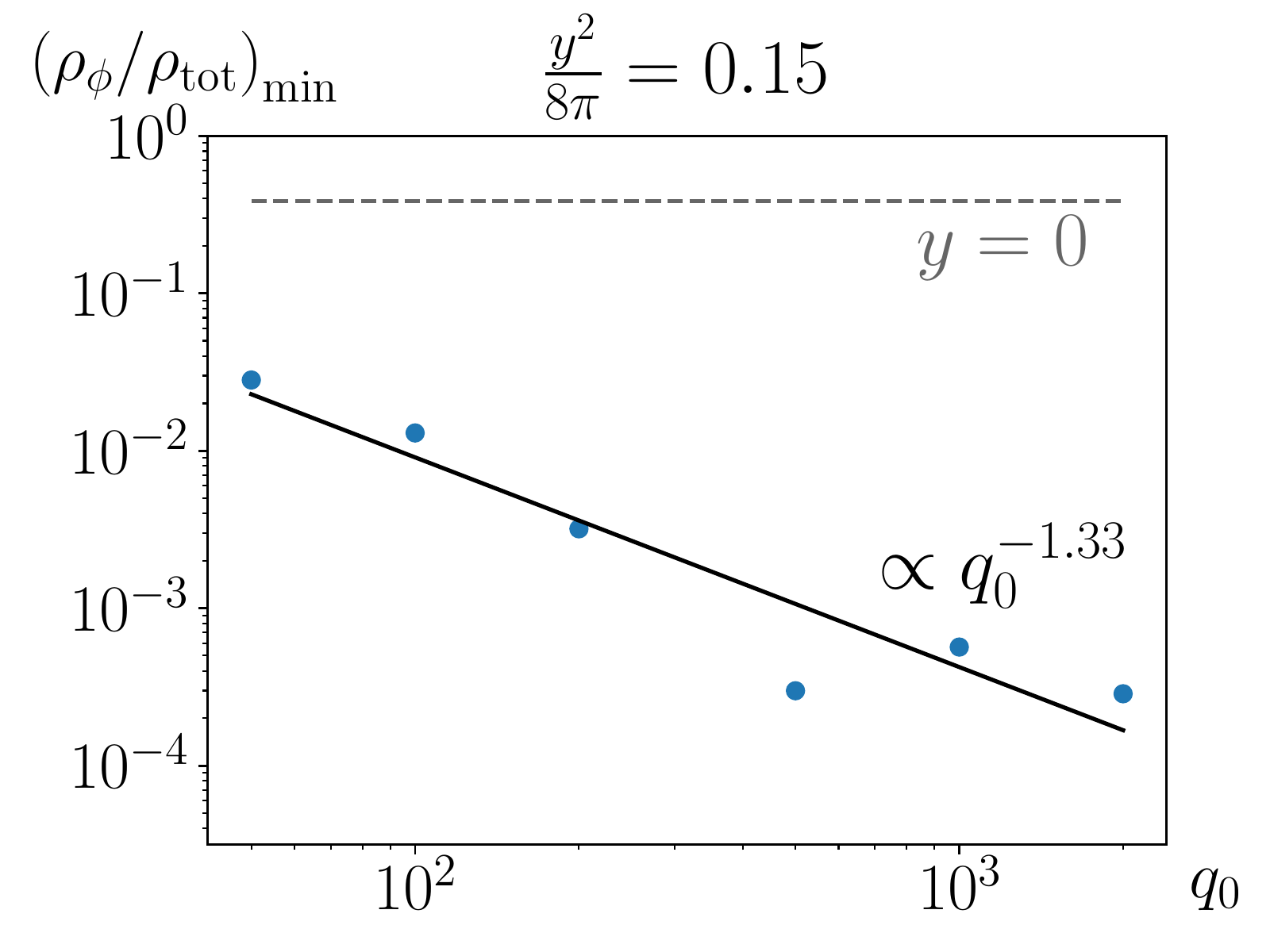}
    \end{subfigure}
    \caption{$(\rho_{\phi}/\rho_{\text{tot}})_{\text{min}}$ as a function of $q_0$ and $y$ for $b = 0.9$, $\lambda=4.05$, $\Phi_0 = f = \MP$. The blue points are the simulation results, and the black line is the best fit with a power law $q_0^{x}$. Each panel also shows in gray the power law best fit for the $y = 0$ case, which is flat at $(\rho_{\phi}/\rho_{\text{tot}})_{\text{min}}\approx 0.5$.}
    \label{fig:qscalingLargeLam}
\end{figure}

We also check if there is a power-law dependence of $(\rho_{\phi}/\rho_{\text{tot}})_{\text{min}}$ on $q_0$, similar to what we found in \Sec{sec:results}. We study the same choices of parameters as \Sec{sec:results}, $y^2/(8\pi) = 0.01$, 0.05, 0.1, and 0.15 for $q_0$ = 50, 100, 200, 500, 1000, and 2000. For a given value of $q_0$, we fix $f = \MP$, $\lambda = 4.05$, and $b = 0.9$, which sets the values of $m$ and $M$. This means that for larger value of $q_0$, $m$ and $M$ are further apart and smaller compared to $f$. For all simulations, we use a lattice with $128^3$ points and $L = 2m^{-1}$. The results are shown in \Fig{fig:qscalingLargeLam}. Again we observe a power law scaling of $(\rho_{\phi}/\rho_{\text{tot}})_{\text{min}}$ with $q_0$, with quantitatively similar features as what we present in \Sec{sec:results}: the exponents have similar values, and the energy transfer efficiency also improves with greater value of $y$. As discussed before, we expect the energy transfer efficiency to deteriorate when $y\ll 1$ or $y\gg 1$, which are beyond the range of our simulations.

In summary, the results in \Sec{subsec:enhanced} and this section show quantitatively similar patterns of energy transfer for two quite different mass hierarchies in spillway preheating. 
The common features they share and the net result of enhanced energy transfer efficiency due to the perturbative decays serving as a spillway are expected to persist in the numerically infeasible parameter space where all conditions in Eq.~\eqref{eq:ideal} are satisfied with a smaller $N_f$ (and larger $\lambda$) than that in \Sec{subsec:enhanced}.

\section{Conclusions and Outlook}
\label{sec:conc}

In this article, we have studied a preheating scenario featuring non-perturbative decays of the inflaton, $\phi$, into a daughter scalar, $\chi$, and a perturbative fermionic decay channel $\chi\rightarrow\psi\psi$. We show that in the cases where the perturbative decays of $\chi$ into fermions become efficient after $\chi$ has been significantly excited by the oscillating $\phi$, but before the backreaction of $\chi$ on $\phi$ kicks in, up to $99.99\%$ of the inflaton energy can be transferred into the daughter species. This new class of preheating scenario is unmatched in terms of energy transfer efficiency. We dub it {\it spillway preheating}.

We employ classical lattice simulations to explore the non-perturbative decays of the $\phi$-condensate into the daughter $\chi$ bosons. To incorporate the inherently quantum perturbative decays of $\chi$ into pairs of fermionic $\psi$ particles, we add a phenomenological friction term to the classical Klein-Gordon equations of motion governing the evolution of $\chi$. The fermions are added to the lattice as a homogeneous radiation fluid, $\rho_{\psi}$. The simulations are carried out in an FRW background, expanding in a self-consistent manner. The evolution of the scale factor was determined by the energies and equations of state of the effective $\phi$, $\chi$ and $\psi$ fluids. The excellent energy conservation, better than one part in a thousand even when the energy budget is dominated by the fermionic fluid, is a strong indication for the validity of our effective description of the theory.

We also provide a parametric understanding of the energy transfer efficiency. We show that the minimum fraction of energy density remaining in the inflaton scales as a simple power law in a parameter $q_0$, which is the ratio of the squared mass scales of the daughter scalar and the inflaton. The larger the mass hierarchy between $\chi$ and $\phi$ is, the more efficient the energy transfer becomes. 

There is much more to be explored in spillway preheating, e.g., 
\begin{itemize}
\item With the computational resources we have, we simulate $q_0$ up to 2000 and show that the depletion of the inflaton energy density could be improved by four orders of magnitude, compared to traditional preheating scenarios. Will the simple power-law scaling we find persist for even larger $q_0$'s and what could be the maximum energy transfer efficiency achievable in the scenario? Could this preheating scenario alone be sufficient to complete the phase transition from inflation to the thermal big bang? 
\item We only consider non-perturbative decays of $\phi$ into $\chi$ due to a tachyonic instability. It would be interesting to see if the results change for resonant instabilities coming from, e.g., $\phi^2\chi^2$ interactions. We leave the investigation of the effects of the form of the inflaton couplings on spillway preheating for future work.
\item What are the effects on the cosmological observables, such as the inflationary observables and gravitational waves? Spillway preheating speeds up the transition to a radiation-dominated state of expansion ($w=1/3$), which can reduce the theoretical uncertainties in $n_{\rm s}$ and $r$ significantly \cite{Lozanov:2016hid,Lozanov:2017hjm,Antusch:2020iyq}. We defer the study of such observational effects for the future.
\item Could this very efficient dissipation mechanism and its variants be applied to solve other interesting problems in particle physics, e.g., solve the cosmological moduli problem~\cite{Giblin:2017wlo} or expand the parameter space of dark photon dark matter~\cite{Agrawal:2018vin, Co:2018lka, Dror:2018pdh, Bastero-Gil:2018uel}? 
\end{itemize}

\section*{Acknowledgments}
We thank Mustafa A. Amin for collaboration in the early stage of the project. We thank Matt Reece, Jean-Samuel Roux and Scott Watson for useful feedback on the manuscript. JF is supported by the DOE grant DE-SC-0010010 and NASA grant 80NSSC18K1010. The work of KL is supported in part by the US Department of Energy through grant DE-SC0015655. QL is supported by the DOE Grant DE-SC-0013607 and the NASA Grant 80NSSC20K0506. 
This research was conducted using computational resources and services at the Center for Computation and Visualization, Brown University.

\bibliography{ref}
\bibliographystyle{utphys}
\end{document}